\documentclass[english,prd,superscriptaddress,nofootinbib,preprintnumbers,notitlepage]{revtex4-1}
\usepackage[T1]{fontenc}
\usepackage[latin1]{inputenc}
\usepackage{bm}
\usepackage{amsmath}
\usepackage{amssymb}
\usepackage{graphicx}
\usepackage{esint}

\makeatletter
\@ifundefined{textcolor}{}
{%
 \definecolor{BLACK}{gray}{0}
 \definecolor{WHITE}{gray}{1}
 \definecolor{RED}{rgb}{1,0,0}
 \definecolor{GREEN}{rgb}{0,1,0}
 \definecolor{BLUE}{rgb}{0,0,1}
 \definecolor{CYAN}{cmyk}{1,0,0,0}
 \definecolor{MAGENTA}{cmyk}{0,1,0,0}
 \definecolor{YELLOW}{cmyk}{0,0,1,0}
}

\newcommand{\tA}{\tilde{A}}
\newcommand{\tB}{\tilde{B}}
\newcommand{\tC}{\tilde{C}}

\usepackage{babel}

\makeatother

\usepackage{babel}
\begin{document}

\title{Cosmological constraints on extended Galileon models}

\author{Antonio De Felice}

\affiliation{TPTP \& NEP, The Institute for Fundamental Study, Naresuan University,
Phitsanulok 65000, Thailand}

\affiliation{Thailand Center of Excellence in Physics, Ministry of Education,
Bangkok 10400, Thailand}

\author{Shinji Tsujikawa}

\affiliation{Department of Physics, Faculty of Science, Tokyo University of Science,
1-3, Kagurazaka, Shinjuku-ku, Tokyo 162-8601, Japan}
\begin{abstract}
The extended Galileon models possess tracker solutions with de Sitter
attractors along which the dark energy equation of state is constant
during the matter-dominated epoch, i.e. $w_{{\rm DE}}=-1-s$, where
$s$ is a positive constant. Even with this phantom equation of state
there are viable parameter spaces in which the ghosts and Laplacian
instabilities are absent. Using the observational data of the supernovae
type Ia, the cosmic microwave background (CMB), and baryon acoustic
oscillations, we place constraints on the tracker solutions at the
background level and find that the parameter $s$ is constrained to
be $s=0.034_{-0.034}^{+0.327}$ (95 \% CL) in the flat Universe. In
order to break the degeneracy between the models we also study the
evolution of cosmological density perturbations relevant to the large-scale
structure (LSS) and the Integrated-Sachs-Wolfe (ISW) effect in CMB.
We show that, depending on the model parameters, the LSS and the ISW
effect is either positively or negatively correlated. It is then possible
to constrain viable parameter spaces further from the observational
data of the ISW-LSS cross-correlation as well as from the matter power
spectrum. 
\end{abstract}

\date{\today}

\maketitle

\section{Introduction}


The main target of the dark energy research over the next few years
or so is to distinguish between the $\Lambda$-Cold-Dark-Matter ($\Lambda$CDM)
model and dynamical models with time-varying equations of state $w_{{\rm DE}}$.
{}From the observational data of WMAP7 combined with baryon acoustic
oscillations (BAO) \cite{Percival} and the Hubble constant measurement
\cite{Riess09}, Komatsu \textit{et al.} \cite{WMAP7} derived the
bound $w_{{\rm DE}}=-1.10\pm0.14$ (68 \% CL) for the constant equation
of state. Adding the supernovae type Ia (SN Ia) data provides tighter
constraints on $w_{{\rm DE}}$, but still the phantom equation of
state ($w_{{\rm DE}}<-1$) is allowed by the joint data analysis \cite{WMAP7}.
This property persists for the time-varying dark energy equation of
state with the parametrization such as $w_{{\rm DE}}=w_{0}+w_{a}(1-a)$
\cite{para}, where $a$ is the scale factor \cite{Wang}.

In the framework of General Relativity (GR) it is generally difficult
to construct theoretically consistent models of dark energy which
realize $w_{{\rm DE}}<-1$. In quintessence \cite{quin} with a slowly
varying scalar-field potential, for example, the field equation of
state is always larger than $-1$. A ghost field with a negative kinetic
energy leads to $w_{{\rm DE}}<-1$ \cite{phantom}, but such a field
is plagued by a catastrophic instability of the vacuum associated
with the spontaneous creation of ghost and photon pairs \cite{Trodden}.

In modified gravitational theories it is possible to realize $w_{{\rm DE}}<-1$
without having ghosts and Laplacian-type instabilities (see Refs.~\cite{review}).
In $f(R)$ gravity, where the Lagrangian $f$ is a function of the
Ricci scalar $R$, the dark energy equation of state crosses the cosmological
constant boundary ($w_{{\rm DE}}=-1$) \cite{Hu,Amen08,Tsuji08,Moto}
for the viable models constructed to satisfy cosmological and local
gravity constraints \cite{Hu,Amen08,fRviable,Tsuji08}. This is also
the case for the Brans-Dicke theory \cite{Brans} with a field potential
which accommodates the chameleon mechanism \cite{chame} to suppress
the propagation of the fifth force \cite{Yoko}. In modified gravity
models of dark energy based on the chameleon mechanism (including
$f(R)$ gravity), the effective potential of a scalar degree of freedom
needs to be carefully designed to pass cosmological and local gravity
constraints \cite{Gan}.

There is another class of modified gravity models of dark energy in
which a nonlinear self-interaction of a scalar degree of freedom $\phi$
can lead to the recovery of GR in a local region through the Vainshtein
mechanism \cite{Vain}. The representative models of this class are
those based on the Dvali-Gabadadze-Porrati (DGP) braneworld \cite{DGP}
and the Galileon gravity \cite{Nicolis} (see Refs.~\cite{DGPnon,Galileonva}
for the implementation of the Vainshtein mechanism in these models).
The nonlinear interaction of the form $(\partial\phi)^{2}\square\phi$,
which appears from the brane-bending mode in the DGP model \cite{DGPnon},
gives rise to the field equation invariant under the Galilean shift
$\partial_{\mu}\phi\to\partial_{\mu}\phi+b_{\mu}$ in the flat spacetime.
This was extended to more general field Lagrangians satisfying the
Galilean symmetry in the limit of the Minkowski spacetime \cite{Nicolis,DEV}.

The cosmology based on the covariant Galileon or on its modified versions
has been studied by many authors \cite{Galileondark,Galileoninf}.
In Refs.~\cite{DePRL,DePRD} the dynamics of dark energy was investigated
in the presence of the full covariant Galileon Lagrangian. In this
model the solutions with different initial conditions converge to
a common trajectory (tracker). Along the tracker the dark energy equation
of state $w_{{\rm DE}}$ changes as $-7/3$ (radiation era) $\to$
$-2$ (matter era) $\to$ $-1$ (de Sitter era) \cite{DePRL,DePRD,Vikman10,Kimura}.
There exists a viable parameter space in which the ghosts and Laplacian
instabilities are absent. However, the joint analysis based on the
observational data of SN Ia, CMB, and BAO shows that the tracker solution
is disfavored because of the large deviation of $w_{{\rm DE}}$ from
$-1$ during the matter era \cite{Nesseris,Kimura}. The solutions
that approach the tracker only at late times are allowed from the
combined data analysis \cite{Nesseris}.

As an extension of the covariant Galileon model, Deffayet \textit{et
al.} \cite{Deffayet11} obtained the most general Lagrangian in scalar-tensor
theories with second-order equations of motion. In four dimensions
the corresponding Lagrangian is of the form (\ref{Lagsum}) with the
four functions (\ref{eachlag2})-(\ref{eachlag5}) given below. In
fact this is equivalent to the Lagrangian found by Horndeski \cite{Horndeski}
more than 3 decades ago \cite{Koba11,Cope}. The conditions for the
avoidance of ghosts and Laplacian instabilities were recently derived
in Ref.~\cite{Defe2011} in the presence of two perfect fluids (non-relativistic
matter and radiation).

The covariant Galileon corresponds to the choice $K=-c_{2}X$, 
$G_{3}=c_{3}X/M^{3}$,
$G_{4}=M_{{\rm pl}}^{2}/2-c_{4}X^{2}/M^{6}$, $G_{5}=3c_{5}X^{2}/M^{9}$
in Eqs.~(\ref{eachlag2})-(\ref{eachlag5}), where $c_{i}$'s are
dimensionless constants, $X=-\partial^{\mu}\phi\partial_{\mu}\phi/2$,
$M_{{\rm pl}}$ is the reduced Planck mass, and $M$ is a constant
having the dimension of mass. Kimura and Yamamoto \cite{Kimura} studied
the model with the functions $K=-c_{2}X$, $G_{3}=c_{3}M^{1-4n}X^{n}$
($n\geq1$), $G_{4}=M_{{\rm pl}}^{2}/2$, and $G_{5}=0$, in which
case the dark energy equation of state during the matter era is given
by $w_{{\rm DE}}=-1-s$ with $s=1/(2n-1)>0$. At the background level
this is equivalent to the Dvali-Turner model \cite{DTurner}, which
can be consistent with the observational data for $n$ larger than
the order of 1. If we consider the evolution of cosmological perturbations,
the LSS tends to be anti-correlated with the late-time ISW effect.
This places the tight bound on the power $n$, as $n>4.2\times10^{3}$
(95\% CL) \cite{Kimura2}, in which case the dark energy equation
of state is practically indistinguishable from that 
in the $\Lambda$CDM model.

In Ref.~\cite{Defe2011} the present authors proposed more general
extended Galileon models with the functions $K=-c_{2}M_{2}^{4(1-p_{2})}X^{p_{2}}$,
$G_{3}=c_{3}M_{3}^{1-4p_{3}}X^{p_{3}}$, $G_{4}=M_{{\rm pl}}^{2}/2-c_{4}M_{4}^{2-4p_{4}}X^{p_{4}}$,
and $G_{5}=3c_{5}M_{5}^{-(1+4p_{5})}X^{p_{5}}$, where the masses
$M_{i}$'s are fixed by the Hubble parameter at the late-time de Sitter
solution with $\dot{\phi}=$\,constant. For the powers $p_{2}=p$,
$p_{3}=p+(2q-1)/2$, $p_{4}=p+2q$, $p_{5}=p+(6q-1)/2$, where $p$
and $q$ are positive constants, there exists a tracker solution characterized
by $H\dot{\phi}^{2q}=$\,constant. During the matter-dominated epoch
one has $w_{{\rm DE}}=-1-s$, where $s=p/(2q)$, along the tracker.
This covers the model of Kimura and Yamamoto \cite{Kimura} as a specific
case ($p=1$, $q=n-1/2$, $c_{4}=0$, $c_{5}=0$). In the presence
of the nonlinear field self-interactions in $G_{4}$ and $G_{5}$,
the degeneracy of the background tracker solution for given values
of $p$ and $q$ is broken by considering the evolution of cosmological
perturbations. Hence the ISW-LSS anti-correlation found in Refs.~\cite{Kimura,Kimura2}
for $c_{4}=c_{5}=0$ should not be necessarily present for the models
with non-zero values of $c_{4}$ and $c_{5}$.

In this paper we first place constraints on the tracker solution in
the extended Galileon models by using the recent observational data
of SN Ia, CMB, and BAO. The bound on the value $s=p/(2q)$ is derived
from the background cosmic expansion history with/without the cosmic
curvature $K$. We then study the evolution of cosmological density
perturbations in the presence of non-relativistic matter to break
the degeneracy of the tracker solution at the background level. We
will show that the LSS and the ISW effect are either positively or
negatively correlated, depending on the parameters $c_{4}$ and $c_{5}$.
This information should be useful to distinguish between the extended
Galileon models with different values of $c_{4}$ and $c_{5}$ from
observations.

\section{Background field equations}
\label{fieldeqsec} 

We start with the following Lagrangian 
\begin{equation}
{\cal L}=\sum_{i=2}^{5}{\cal L}_{i}\,,\label{Lagsum}
\end{equation}
 where 
\begin{eqnarray}
{\cal L}_{2} & = & K(X),\label{eachlag2}\\
{\cal L}_{3} & = & -G_{3}(X)\Box\phi,\\
{\cal L}_{4} & = & G_{4}(X)\, R+G_{4,X}\,[(\Box\phi)^{2}-(\nabla_{\mu}\nabla_{\nu}\phi)\,(\nabla^{\mu}\nabla^{\nu}\phi)]\,,\\
{\cal L}_{5} & = & G_{5}(X)\, G_{\mu\nu}\,(\nabla^{\mu}\nabla^{\nu}\phi)-(G_{5,X}/6)[(\Box\phi)^{3}-3(\Box\phi)\,(\nabla_{\mu}\nabla_{\nu}\phi)\,(\nabla^{\mu}\nabla^{\nu}\phi)+2(\nabla^{\mu}\nabla_{\alpha}\phi)\,(\nabla^{\alpha}\nabla_{\beta}\phi)\,(\nabla^{\beta}\nabla_{\mu}\phi)]\,.\label{eachlag5}
\end{eqnarray}
 $K$ and $G_{i}$ ($i=3,4,5$) are functions in terms of the field
kinetic energy $X=-\partial^{\mu}\phi\partial_{\mu}\phi/2$ with $G_{i,X}\equiv dG_{i}/dX$,
$R$ is the Ricci scalar, and $G_{\mu\nu}$ is the Einstein tensor.
If we allow the $\phi$-dependence for the functions $K$ and $G_{i}$
as well, the Lagrangian (\ref{Lagsum}) corresponds to the most general
Lagrangian in scalar-tensor theories \cite{Horndeski,Deffayet11}.
In order to discuss models relevant to dark energy we also take into
account the perfect fluids of non-relativistic matter and radiation
(with the Lagrangians ${\cal L}_{m}$ and ${\cal L}_{r}$ respectively),
in which case the total 4-dimensional action is given by 
\begin{equation}
S=\int d^{4}x\sqrt{-g}({\cal L}+{\cal L}_{m}+{\cal L}_{r})\,.\label{action}
\end{equation}
 In the following we focus on the extended Galileon models \cite{Defe2011}
in which $K$ and $G_{i}$ are given by 
\begin{equation}
K=-c_{2}M_{2}^{4(1-p_{2})}X^{p_{2}}\,,\qquad G_{3}=c_{3}M_{3}^{1-4p_{3}}X^{p_{3}}\,,\qquad G_{4}=M_{{\rm pl}}^{2}/2-c_{4}M_{4}^{2-4p_{4}}X^{p_{4}}\,,\qquad G_{5}=3c_{5}M_{5}^{-(1+4p_{5})}X^{p_{5}},\label{geGali}
\end{equation}
 where $M_{{\rm pl}}$ is the reduced Planck mass, $c_{i}$ and $p_{i}$
($i=2,3,4,5$) are dimensionless constants, and $M_{i}$ ($i=2,3,4,5$)
are constants having the dimension of mass. In the flat Universe it
was shown in Ref.~\cite{Defe2011} that tracker solutions characterized
by the condition $H\dot{\phi}^{2q}={\rm constant}$ ($q>0$ and a
dot represents a derivative with respect to cosmic time $t$) are
present for 
\begin{equation}
p_{2}=p\,,\qquad p_{3}=p+(2q-1)/2\,,\qquad p_{4}=p+2q\,,\qquad p_{5}=p+(6q-1)/2\,.\label{power}
\end{equation}
 The covariant Galileon \cite{DEV} corresponds to $p=1$ and $q=1/2$,
i.e. $p_{2}=p_{3}=1$, $p_{4}=p_{5}=2$.

We will extend the analysis to the general Friedmann-Lema\^{i}tre-Robertson-Walker
(FLRW) background with the cosmic curvature $K$: 
\begin{equation}
{\rm d}s^{2}=-{\rm d}t^{2}+a^{2}(t)\left[\frac{{\rm d}r^{2}}{1-Kr^{2}}+r^{2}({\rm d}\theta^{2}+\sin^{2}\theta\,{\rm d}\phi^{2})\right]\,,\label{metric}
\end{equation}
 where $a(t)$ is the scale factor. The closed, flat, and open geometries
correspond to $K>0$, $K=0$, and $K<0$, respectively. For the theories
given by the action (\ref{action}) the dynamical equations of motion
are 
\begin{eqnarray}
 &  & 3H^{2}M_{{\rm pl}}^{2}=\rho_{{\rm DE}}+\rho_{m}+\rho_{r}+\rho_{K}\,,\label{be1}\\
 &  & (3H^{2}+2\dot{H})M_{{\rm pl}}^{2}=-P_{{\rm DE}}-\rho_{r}/3+\rho_{K}/3\,,\label{be2}\\
 &  & \dot{\rho}_{m}+3H\rho_{m}=0\,,\\
 &  & \dot{\rho}_{r}+4H\rho_{r}=0\,,\\
 &  & \dot{\rho}_{K}+2H\rho_{K}=0\,.
\end{eqnarray}
 Here $H\equiv\dot{a}/a$, $\rho_{K}\equiv-3KM_{{\rm pl}}^{2}/a^{2}$,
$\rho_{m}$ and $\rho_{r}$ are the energy densities of non-relativistic
matter and radiation respectively, and 
\begin{eqnarray}
\rho_{{\rm DE}} & \equiv & 2XK_{,X}-K+6H\dot{\phi}XG_{3,X}-6H^{2}\tilde{G}_{4}+24H^{2}X(G_{4,X}+XG_{4,XX})+2H^{3}\dot{\phi}X(5G_{5,X}+2XG_{5,XX})\,,\\
P_{{\rm DE}} & \equiv & K-2X\ddot{\phi}\, G_{3,X}+2(3H^{2}+2\dot{H})\tilde{G}_{4}-4(3H^{2}X+H\dot{X}+2\dot{H}X)G_{4,X}-8HX\dot{X}G_{4,XX}\nonumber \\
 &  & -2X(2H^{3}\dot{\phi}+2H\dot{H}\dot{\phi}+3H^{2}\ddot{\phi})G_{5,X}-4H^{2}X^{2}\ddot{\phi}\, G_{5,XX}\,,
\end{eqnarray}
 where $\tilde{G}_{4}\equiv G_{4}-M_{{\rm pl}}^{2}/2=-c_{4}M_{4}^{2-4p_{4}}X^{p_{4}}$.

From Eqs.~(\ref{be1}) and (\ref{be2}) we find that there exists
a de Sitter solution characterized by $\dot{H}=0$ and $\ddot{\phi}=0$.
In order to discuss the cosmological dynamics we introduce the dimensionless
variables \cite{Defe2011} 
\begin{equation}
r_{1}\equiv\left(\frac{x_{{\rm dS}}}{x}\right)^{2q}\left(\frac{H_{{\rm dS}}}{H}\right)^{1+2q}\,,\qquad r_{2}\equiv\left[\left(\frac{x}{x_{{\rm dS}}}\right)^{2}\frac{1}{r_{1}^{3}}\right]^{\frac{p+2q}{1+2q}}\,,\qquad\Omega_{r}\equiv\frac{\rho_{r}}{3H^{2}M_{{\rm pl}}^{2}}\,,\label{eq:defr1r2}
\end{equation}
 where $x\equiv\dot{\phi}/(HM_{{\rm pl}})$, and the subscript ``dS''
represents the quantities at the de Sitter solution. We relate the
masses $M_{i}$ ($i=2,\cdots,5$) in Eq.~(\ref{geGali}) with $H_{{\rm dS}}$,
as $M_{2}\equiv(H_{{\rm dS}}M_{{\rm pl}})^{1/2}$, $M_{3}\equiv(H_{{\rm dS}}^{-2p_{3}}M_{{\rm pl}}^{1-2p_{3}})^{1/(1-4p_{3})}$,
$M_{4}\equiv(H_{{\rm dS}}^{-2p_{4}}M_{{\rm pl}}^{2-2p_{4}})^{1/(2-4p_{4})}$,
and $M_{5}\equiv(H_{{\rm dS}}^{2+2p_{5}}M_{{\rm pl}}^{2p_{5}-1})^{1/(1+4p_{5})}$.
The existence of de Sitter solutions demands that the coefficients
$c_{2}$ and $c_{3}$ are related with $c_{4}$ and $c_{5}$, via
\begin{equation}
c_{2}=\frac{3}{2}\left(\frac{2}{x_{{\rm dS}}^{2}}\right)^{p}(3\alpha-4\beta+2)\,,\qquad c_{3}=\frac{\sqrt{2}}{2p+q-1}\left(\frac{2}{x_{{\rm dS}}^{2}}\right)^{p+q}\left[3(p+q)(\alpha-\beta)+p\right]\,,\label{c3}
\end{equation}
 where 
\begin{equation}
\alpha\equiv\frac{4(2p_{4}-1)}{3}\left(\frac{x_{{\rm dS}}^{2}}{2}\right)^{p_{4}}c_{4}\,,\qquad\beta\equiv2\sqrt{2}\, p_{5}\left(\frac{x_{{\rm dS}}^{2}}{2}\right)^{p_{5}+1/2}c_{5}\,.\label{ab}
\end{equation}
 The density parameter of dark energy, $\Omega_{{\rm DE}}\equiv\rho_{{\rm DE}}/(3H^{2}M_{{\rm pl}}^{2})$,
can be expressed as 
\begin{equation}
\Omega_{{\rm DE}}=\frac{r_{1}^{\frac{p-1}{2q+1}}r_{2}}{2}\Bigl[r_{1}\left\{ r_{1}\bigl[12(\alpha-\beta)(p+q)+4p-r_{1}(2p-1)(3\alpha-4\beta+2)\bigr]-3\alpha(2p+4q+1)\right\} +4\beta(p+3q+1)\Bigr].\label{eq:OMDE}
\end{equation}
 {}From Eq.~(\ref{be1}) it follows that $\Omega_{{\rm DE}}+\Omega_{m}+\Omega_{r}+\Omega_{K}=1$,
where $\Omega_{m}\equiv\rho_{m}/(3H^{2}M_{{\rm pl}}^{2})$ and $\Omega_{K}\equiv\rho_{K}/(3H^{2}M_{{\rm pl}}^{2})$.

The autonomous equations for $r_{1}$, $r_{2}$, and $\Omega_{r}$
are written in terms of $r_{1}$, $r_{2}$, $\Omega_{r}$, $\alpha$,
$\beta$, $p$, $q$. As in the case of the flat Universe \cite{Defe2011}
one can show that there is a fixed point for the differential equation
of $r_{1}$ characterized by 
\begin{equation}
r_{1}=1\,,
\end{equation}
 {}From the definition of $r_{1}$ in Eq.~(\ref{eq:defr1r2}) this
corresponds to the tracker solution where $H\dot{\phi}^{2q}=$\,constant.
Along the tracker the autonomous equations for $r_{2}$ and $\Omega_{r}$
are 
\begin{eqnarray}
r_{2}' & = & \frac{(p+2q)(\Omega_{r}+3-3r_{2}-\Omega_{K})}{pr_{2}+2q}r_{2}\,,\\
\Omega_{r}' & = & \frac{2q(\Omega_{r}-1-3r_{2}-\Omega_{K})-4pr_{2}}{pr_{2}+2q}\Omega_{r}\,,\label{Omereq}
\end{eqnarray}
 where a prime represents the derivative with respect to $N=\ln a$.
Combining these equations, we obtain the integrated solution 
\begin{equation}
r_{2}=c_{1}a^{4(1+s)}\Omega_{r}^{1+s}\,,\qquad s=\frac{p}{2q}\,,
\end{equation}
 where $c_{1}$ is a constant. For the theoretical consistency the
parameter $s$ is positive \cite{Defe2011}. Since $r_{2}\propto H^{-2(1+s)}$,
the quantity $r_{2}$ grows toward the value 1 at the de Sitter solution.
Along the tracker the density parameter (\ref{eq:OMDE}) is given
by 
\begin{equation}
\Omega_{{\rm DE}}=r_{2}=\frac{1-\Omega_{m,0}-\Omega_{r,0}-\Omega_{K,0}}{\Omega_{r,0}^{1+s}}\, e^{4(1+s)N}\,\Omega_{r}^{1+s}\,,\label{eq:r2Omr}
\end{equation}
 where the subscripts ``0'' represent the values today (the scale
factor $a_{0}=1$, i.e. $N_{0}=\ln a_{0}=0$). Using the relation
$\Omega_{K}/\Omega_{r}=(\Omega_{K,0}/\Omega_{r,0})e^{2N}$ as well,
Eq.~(\ref{Omereq}) reads 
\begin{equation}
\Omega'_{r}=-\frac{1-\Omega_{r}+\Omega_{K,0}e^{2N}\,\Omega_{r}/\Omega_{r,0}+(1-\Omega_{m,0}-\Omega_{r,0}-\Omega_{K,0})(3+4s)\, e^{4(1+s)N}\Omega_{r}^{1+s}/\Omega_{r,0}^{1+s}}{1+(1-\Omega_{m,0}-\Omega_{r,0}-\Omega_{K,0})se^{4(1+s)N}\Omega_{r}^{1+s}/\Omega_{r,0}^{1+s}}\Omega_{r}\,.\label{eq:OmrN}
\end{equation}
 This equation can be solved as 
\begin{equation}
\frac{1-\Omega_{m,0}-\Omega_{r,0}-\Omega_{K,0}}{\Omega_{r,0}^{1+s}}\, e^{4(1+s)N}\,\Omega_{r}^{1+s}+\frac{\Omega_{m,0}}{\Omega_{r,0}}\, e^{N}\Omega_{r}+\Omega_{r}+\frac{\Omega_{K,0}}{\Omega_{r,0}}\, e^{2N}\,\Omega_{r}=1\,,\label{eq:exacts}
\end{equation}
 which is nothing but the relation $\Omega_{{\rm DE}}+\Omega_{m}+\Omega_{r}+\Omega_{K}=1$.
{}From Eq.~(\ref{eq:r2Omr}) the dark energy density parameter evolves
as $\Omega_{{\rm DE}}\propto H^{-2(1+s)}$ and hence 
\begin{equation}
\frac{H}{H_{0}}=\left(\frac{\Omega_{{\rm DE},0}}{\Omega_{{\rm DE}}}\right)^{1/[2(1+s)]}\,.\label{eq:Hr2}
\end{equation}
 Since it is not generally possible to solve Eq.~(\ref{eq:exacts})
for $\Omega_{r}$ in terms of $N$ (apart from some specific
values of $s$ such as $s=1$), we numerically integrate Eq.~(\ref{eq:OmrN})
and find the expression of $H/H_{0}$ by using Eqs.~(\ref{eq:r2Omr})
and (\ref{eq:Hr2}).

\begin{figure}
\begin{centering}
\includegraphics[width=3.3in,height=3.3in]{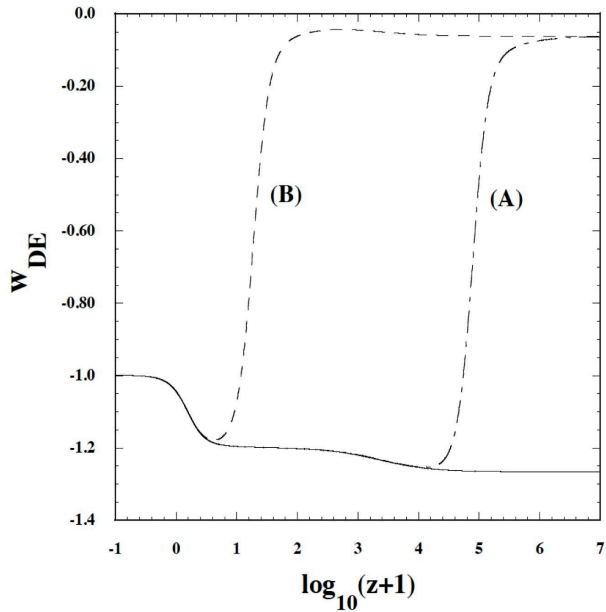} 
\par\end{centering}

\caption{Evolution of $w_{{\rm DE}}$ versus the redshift $z$ for the open
Universe with $\Omega_{K,0}=0.1$ for $p=1$, $q=5/2$, $\alpha=3$,
$\beta=1.45$. The solid curve corresponds to the tracker solution
with the initial conditions $r_{1}=1$, $r_{2}=1.0\times10^{-30}$,
$\Omega_{r}=0.99987$, $\Omega_{K}=4.0\times10^{-12}$ at $\log_{10}(z+1)=7.245$.
In the cases (A) and (B) the initial conditions are chosen to be (A)
$r_{1}=1.0\times10^{-2}$, $r_{2}=1.0\times10^{-23}$, $\Omega_{r}=0.99985$,
$\Omega_{K}=5.0\times10^{-12}$ at $\log_{10}(z+1)=7.210$, and (B)
$r_{1}=3.0\times10^{-6}$, $r_{2}=1.0\times10^{-10}$, $\Omega_{r}=0.9998$,
$\Omega_{K}=1.15\times10^{-11}$ at $\log_{10}(z+1)=6.967$, respectively.}

\centering{}\label{wdefig} 
\end{figure}


Along the tracker the dark energy equation of state $w_{{\rm DE}}\equiv P_{{\rm DE}}/\rho_{{\rm DE}}$
and the effective equation of state $w_{{\rm eff}}=-1-2\dot{H}/(3H^{2})$
are given by 
\begin{equation}
w_{{\rm DE}}=-\frac{3+s(3+\Omega_{r}-\Omega_{K})}{3(1+sr_{2})}\,,\qquad w_{{\rm eff}}=-\frac{r_{2}(3s+3-s\Omega_{K})-\Omega_{r}}{3(1+sr_{2})}\,.\label{wde}
\end{equation}
 In the early cosmological epoch ($r_{2}\ll1$) these reduce to $w_{{\rm DE}}\simeq-1-s(3+\Omega_{r}-\Omega_{K})/3$
and $w_{{\rm eff}}\simeq\Omega_{r}/3$. During the matter era in which
$\{\Omega_{r},|\Omega_{K}|\}\ll1$ it follows that $w_{{\rm DE}}\simeq-1-s<-1$
(for $s>0$) and $w_{{\rm eff}}\simeq0$. At the de Sitter fixed point
($r_{2}=1$) with $\Omega_{r}=\Omega_{K}=0$ one has $w_{{\rm DE}}=w_{{\rm eff}}=-1$.
In Fig.~\ref{wdefig} we plot the evolution of $w_{{\rm DE}}$ versus
the redshift $z=a_{0}/a-1$ in the open Universe with $\Omega_{K,0}=0.1$
for the model parameters $p=1$, $q=5/2$, $\alpha=3$, $\beta=1.45$
(i.e. $s=0.2$). The tracker is shown as a solid curve, along which
$w_{{\rm DE}}$ changes as $-1.267$ (radiation era) $\to$ $-1.2$
(matter era) $\to$ $-1$ (de Sitter era). The effect of the cosmic
curvature $\Omega_{K}$ becomes important only for the late Universe,
which affects the luminosity distance in the SN Ia observations.

If the solutions start from the regime $r_{1}\ll1$, the evolution
of $w_{{\rm DE}}$ is different from Eq.~(\ref{wde}) before they
reach the tracker. For $r_{1}\ll1$ and $r_{2}\ll1$, $w_{{\rm DE}}$
and $w_{{\rm eff}}$ are approximately given by 
\begin{equation}
w_{{\rm DE}}\simeq-\frac{1+\Omega_{r}-\Omega_{K}}{2(2p+6q-1)}\,,\qquad w_{{\rm eff}}\simeq\frac{1}{3}\Omega_{r}\,.\label{wde2}
\end{equation}
 In Fig.~\ref{wdefig} we show the variation of $w_{{\rm DE}}$ for
$p=1$, $q=5/2$, $\alpha=3$, $\beta=1.45$ with two different initial
conditions satisfying $r_{1}\ll1$. In both cases the density parameter
$\Omega_{K}$ today is $\Omega_{K,0}=0.1$. The cases (A) and (B)
correspond to the early and late trackings, respectively. For smaller
initial values of $r_{1}$ the tracking occurs later. As estimated
by Eq.~(\ref{wde2}), $w_{{\rm DE}}$ starts from the value $w_{{\rm DE}}\simeq-1/16$
in the deep radiation era. If the solutions do not reach the tracker
during the matter era (as in the case (B) in Fig.~\ref{wdefig}),
$w_{{\rm DE}}$ temporally approaches the value $-1/32$.

In Ref.~\cite{Nesseris} it was shown that the tracker for the covariant
Galileon ($s=1$) is disfavored from observations, but the late-time
tracking solution is allowed from the data. This property comes from
the fact that for the late-time tracker the deviation of $w_{{\rm DE}}$
from $-1$ is not significant. For $s\ll1$ even the tracker is expected
to be allowed from observations. In such cases the solutions starting
from the initial conditions with $r_{1}\ll1$ should be also compatible
with the data (because even for $s=1$ the late-time tracking solution
is allowed). In the following sections we will focus on the tracker
solution to discuss the background observational constraints and the
evolution of cosmological perturbations.


\section{Observational constraints on the extended Galileon models \label{obsercon} }


In this section we place observational constraints on the tracker
solution from the background cosmic expansion history. We use three
data sets: 1) the CMB shift parameters (WMAP7) \cite{WMAP7}; 2) the
BAO (SDSS7) \cite{Percival}; 3) and the SN Ia (Constitution) \cite{hicken}.
The total chi-square $\chi_{{\rm tot}}^{2}$ for all three combined
data sets will be calculated on a grid representing a chosen set of
available parameters. We then find the minimum on this grid, and consequently
find the 1$\sigma$ and 2$\sigma$ contours.

In order to integrate Eq.~(\ref{eq:OmrN}), once a set of model parameters
(in this case not only $s$, but also $\Omega_{r,0}$, $\Omega_{m,0}$,
and $\Omega_{K,0}$) is given, we have the choice of one initial condition,
that is $\Omega_{r,i}\equiv\Omega_{r}(N_{i})$. In principle it is
possible to solve Eq.~(\ref{eq:OmrN}) backwards for a given value
of $\Omega_{r}(0)=\Omega_{r,0}$, but we find that the integrated
results are prone to numerical instabilities. Therefore, it is more
convenient to obtain the expression of $\Omega_{r,i}$ for a given
$N_{i}<0$ (chosen to be $N_{i}=-\ln(1+z_{i})$, where the initial
redshift is $z_{i}=1.76\times10^{7}$), which gives $\Omega_{r}(N=0)=\Omega_{r,0}$
after solving the differential equation.

For given $\Omega_{r,0}$ and the other parameters we can use Eq.~(\ref{eq:exacts})
to obtain the desired value of $\Omega_{r,i}$. At $N=N_{i}$, Eq.~(\ref{eq:exacts})
is written as 
\begin{equation}
\left(1+\frac{\Omega_{K,0}}{\Omega_{r,0}}\, e^{2N_{i}}+\frac{\Omega_{m,0}}{\Omega_{r,0}}\, e^{N_{i}}\right)\Omega_{r,i}-1=-\frac{1-\Omega_{m,0}-\Omega_{K,0}-\Omega_{r,0}}{\Omega_{r,0}^{1+s}}\, e^{4(1+s)N_{i}}\,\Omega_{r,i}^{1+s}\,.\label{eq:exa2-1}
\end{equation}
 Since $\Omega_{r,i}\approx1$ during the radiation domination, it
is possible to solve this equation iteratively by assuming that the
l.h.s. is a small correction (indeed the r.h.s. corresponds to $-\Omega_{{\rm DE},i}$).
At 0-th order the solution of Eq.~(\ref{eq:exa2-1}) is given by
\begin{equation}
\Omega_{r,i}^{(0)}=\frac{1}{1+(\Omega_{K,0}/\Omega_{r,0})\, e^{2N_{i}}+(\Omega_{m,0}/\Omega_{r,0})\, e^{N_{i}}}\,.
\end{equation}
 At first order we find 
\begin{equation}
\Omega_{r,i}^{(1)}=\frac{1-(1-\Omega_{m,0}-\Omega_{K,0}-\Omega_{r,0})e^{4(1+s)N_{i}}\,[\Omega_{r,i}^{(0)}/\Omega_{r,0}]^{1+s}}{1+(\Omega_{K,0}/\Omega_{r,0})\, e^{2N_{i}}+(\Omega_{m,0}/\Omega_{r,0})\, e^{N_{i}}}\,.
\end{equation}
 This process can be iterated up to the desired precision. In the
numerical code, we employ the solution derived after the three iterations,
that is $\Omega_{r,i}\simeq\Omega_{r,i}^{(3)}$. Since the late-time
de Sitter background is an attractor, small differences in the initial
conditions do not lead to very different final solutions. Therefore
we indeed find that the three iterations are sufficient to derive
the parameter $\Omega_{r,0}$ accurately.

In the following we first discuss the method for carrying out the
likelihood analysis to confront the tracker solution with observations
and then proceed to constrain the model parameters.

\subsection{CMB shift parameters}

We use the data of the CMB shift parameters provided by WMAP7, which
are related with the positions of the CMB acoustic peaks. These quantities
are affected by the cosmic expansion history from the decoupling epoch
to today. The redshift at the decoupling is known by means of the
fitting formula of Hu and Sugiyama \cite{HuSugi} 
\begin{equation}
z_{*}=1048\,[1+0.00124(\Omega_{b,0}h^{2})^{-0.738}]\,[1+g_{1}\,(\Omega_{m,0}h^{2})^{g_{2}}]\,,
\end{equation}
 where $g_{1}=0.0783\,(\Omega_{b,0}h^{2})^{-0.238}/[1+39.5\,(\Omega_{b,0}h^{2})^{0.763}]$,
$g_{2}=0.560/[1+21.1\,(\Omega_{b,0}h^{2})^{1.81}]$, 
$h=H_{0}/[100\,\,{\rm km\, sec^{-1}\, Mpc^{-1}}]$,
and $\Omega_{b,0}$ corresponds to the today's density parameter of
baryons. The shift of the CMB acoustic peaks can be quantified by
the two shift parameters \cite{bond} 
\begin{equation}
{\cal R}=\sqrt{\frac{\Omega_{m,0}}{\Omega_{K,0}}}\sinh\left(\sqrt{\Omega_{K,0}}\int_{0}^{z_{*}}\frac{dz}{H(z)/H_{0}}\right),\qquad\quad l_{a}=\frac{\pi\, d_{a}^{(c)}(z_{*})}{r_{s}(z_{*})}\,,
\end{equation}
 where $r_{s}(z_{*})$ corresponds to the sound horizon at the decoupling,
given by 
\begin{equation}
r_{s}(z_{*})=\int_{z_{*}}^{\infty}\frac{dz}{H(z)\,\sqrt{3\{1+3\Omega_{b,0}/[4\Omega_{\gamma,0}(1+z)]\}}}\,.
\end{equation}
 Note that $\Omega_{\gamma,0}$ is the today's value of photon energy
density and $d_{a}^{(c)}(z_{*})$ is the comoving angular distance
to the last scattering surface defined by $d_{a}^{(c)}(z_{*})={\cal R}/[H_{0}\sqrt{\Omega_{m,0}}]$.

The likelihood values of $l_{a},{\cal R},z_{*}$ provided
by the WMAP7 data \cite{WMAP7} are $l_{a}=302.09\pm0.76$, ${\cal R}=1.725\pm0.018$,
and $z_{*}=1091.3\pm0.91$. The chi-square associated with this measurement
is 
\begin{equation}
\chi_{{\rm CMB}}^{2}=(l_{a}-302.09,{\cal R}-1.725,z_{*}-1091.3)\,\bm{C}_{{\rm CMB}}^{-1}\left(\begin{array}{c}
l_{a}-302.09\\
{\cal R}-1.725\\
z_{*}-1091.3
\end{array}\right)\,,
\end{equation}
 where the inverse covariance matrix is 
\begin{equation}
\bm{C}_{{\rm CMB}}^{-1}=\left(\begin{array}{ccc}
2.305 & 29.698 & -1.333\\
29.698 & 6825.27 & -113.18\\
-1.333 & -113.18 & 3.414
\end{array}\right)\,.
\end{equation}

\subsection{BAO}

We also employ the data of BAO measured by the SDSS7 \cite{Percival}.
The redshift $z_{d}$ at which the baryons are released from the Compton
drag of photons is given by the fitting formula of Eisenstein and
Hu \cite{eisen}: 
\begin{equation}
z_{d}=\frac{1291\,(\Omega_{m,0}h^{2})^{0.251}}{1+0.659\,(\Omega_{m,0}h^{2})^{0.828}}\,[1+b_{1}\,(\Omega_{b,0}h^{2})^{b_{2}}]\,,
\end{equation}
 where $b_{1}=0.313\,(\Omega_{m,0}h^{2})^{-0.419}[1+0.607\,(\Omega_{m,0}h^{2})^{0.674}]$
and $b_{2}=0.238(\Omega_{m,0}h^{2})^{0.223}$. We define the effective
BAO distance 
\begin{equation}
D_{V}(z)=\left[d_{A}^{2}(z)\,(1+z)^{2}z/H(z)\right]^{1/3}\,,
\end{equation}
 where $d_{A}(z)$ is the diameter distance given by 
\begin{equation}
d_{A}(z)=\frac{1}{1+z}\,\frac{1}{H_{0}\sqrt{\Omega_{K,0}}}\sinh\!\left[\sqrt{\Omega_{K,0}}\int_{0}^{z}\frac{d\tilde{z}}{H(\tilde{z})/H_{0}}\right]\,.
\end{equation}

The BAO observations constrain the ratio $r_{s}(z_{d})/D_{V}(z)$
at particular redshifts $z$, where $r_{s}(z_{d})$ is the sound horizon
for $z=z_{d}$. At $z=0.2$ and $z=0.35$ the recent observational
bounds are $r_{s}(z_{d})/D_{V}(0.2)=0.1905\pm0.0061$ and $r_{s}(z_{d})/D_{V}(0.35)=0.1097\pm0.0036$.
The chi-square associated with the BAO is evaluated as 
\begin{equation}
\chi_{{\rm BAO}}^{2}=\left(r_{s}(z_{d})/D_{V}(0.2)-0.1905,r_{s}(z_{d})/D_{V}(0.35)-0.1097\right)\bm{C}_{{\rm BAO}}^{-1}\left(\begin{array}{c}
r_{s}(z_{d})/D_{V}(0.2)-0.1905\\
r_{s}(z_{d})/D_{V}(0.35)-0.1097
\end{array}\right)\,,
\end{equation}
 where the inverse covariance matrix is \cite{Percival} 
\begin{equation}
\bm{C}_{{\rm BAO}}^{-1}=\left(\begin{array}{cc}
30124 & -17227\\
-17227 & 86977
\end{array}\right)\,.
\end{equation}

\subsection{SN Ia}

Finally we consider the experimental bounds coming from the observations
of the SN Ia standard candles. The apparent magnitudes, together with
their absolute magnitudes, can be used to generate the following chi-square
\cite{laz} 
\begin{equation}
\chi_{{\rm SN\, Ia}}^{2}=\sum_{i}\frac{\mu_{{\rm obs}}(z_{i})-\mu_{{\rm th}}(z_{i})}{\sigma_{\mu,i}^{2}}\,,
\end{equation}
 where $\sigma_{\mu,i}^{2}$ are the errors on the data, and $\mu_{{\rm th}}$
is the theoretical distance modulus defined as 
\begin{equation}
\mu_{{\rm th}}(z_{i})=5\log_{10}\bigl[\bar{d}_{L}(z_{i})\bigr]+\mu_{0}\,.
\end{equation}
 Here $\bar{d}_{L}(z)$ and $\mu_{0}$ are given, respectively, by
\begin{equation}
\bar{d}_{L}(z)=(1+z)^{2}\, H_{0}\, d_{A}(z)\,,\qquad\mu_{0}=42.38-5\log_{10}h\,.
\end{equation}
 In the following we will make use of the Constitution SN Ia data
sets provided in Ref.~\cite{hicken} (see also Ref.~\cite{amanulla}).

\subsection{Observational constraints on the tracker}

We now define the total chi-square as 
\begin{equation}
\chi^{2}=\chi_{{\rm CMB}}^{2}+\chi_{{\rm BAO}}^{2}+\chi_{{\rm SN\, Ia}}^{2}\,.
\end{equation}
 In the following we reduce the numerical complexity by fixing several
parameters, as $h=0.71$, $\Omega_{b,0}=0.02258h^{2}$, $\Omega_{\gamma,0}=2.469\times10^{-5}h^{-2}$,
and $\Omega_{r,0}=\Omega_{\gamma,0}(1+0.2271\,{\cal N}_{{\rm eff}})$
\cite{WMAP7}, where the relativistic degrees of freedom are set to
be ${\cal N}_{{\rm eff}}=3.04$. Then two analysis will be performed
for the tracker solution: 1) the flat case, $\Omega_{K,0}=0$, for
which two free parameters, $\Omega_{m,0}$ and $s$, are left to be
varied; 2) the non-flat case, for which the additional free parameter,
$\Omega_{K,0}$, is varied as well.

Later on, when we compare models with different number of free parameters,
we will also make use of the Akaike Information Criterion (AIC) method
(see e.g.,\ \cite{liddle}). For each model the AIC is defined as
\begin{equation}
{\rm AIC}=\chi_{{\rm min}}^{2}+2{\cal P}\,,
\end{equation}
 where ${\cal P}$ is the number of free parameters in the model,
and $\chi_{{\rm min}}^{2}$ is the minimum value for $\chi^{2}$ in
the chosen parameter space. The smaller the AIC the better the model.
To be more precise, if the difference of $\chi^{2}$ between two different
models is in the range $0<\Delta({\rm AIC})<2$, the models are considered
to be equivalent, whereas if $\Delta({\rm AIC})>2$, the data prefer
one model with respect to the other.

\subsubsection{Flat case: $\Omega_{K,0}=0$}

In this case we compute the $\chi^{2}$ on a grid in the intervals
$0\leq s<0.9$ and $0.25<\Omega_{m,0}<0.32$. The minimum value of
$\chi^{2}$ is found to be $\chi_{{\rm min}}^{2}=468.876$ for the
model parameters $s=0.03446$ and $\Omega_{m,0}=0.27159$. Then we
calculate the difference of $\chi^{2}$ at each grid point, that is,
$\Delta\chi^{2}=\chi^{2}-\chi_{{\rm min}}^{2}$. When $\Delta\chi^{2}\geq2.88$
the chi-square distribution, with two free parameters, excludes the
models with those values of $\chi^{2}$ at 68\% confidence level (1$\sigma$),
whereas when $\Delta\chi^{2}\geq5.99$ those models are excluded at
95\% CL (2$\sigma$).

\begin{figure}
\begin{centering}
\includegraphics[width=4.5in,height=3.3in]{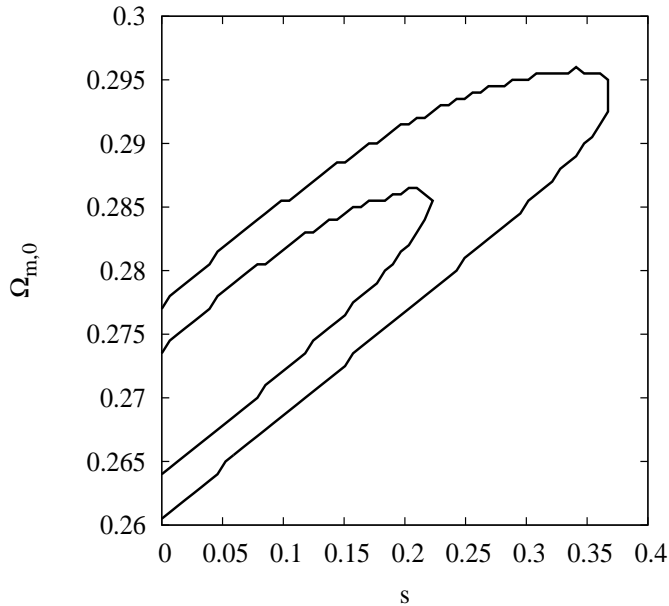} 
\par\end{centering}

\caption{ Observational contours for the tracker solution in the ($s,\Omega_{m,0}$)
plane for the flat Universe ($\Omega_{K,0}=0$). 1$\sigma$ and 2$\sigma$
contours correspond to the internal and external lines, respectively.}

\centering{}\label{Obs-1} 
\end{figure}

Our numerical results are plotted in Fig.~\ref{Obs-1}. Even if we
use the Gaussian likelihood function $P\propto e^{-\chi^{2}/2}$,
we find that the observational contours are similar to those given
in Fig.~\ref{Obs-1}. The parameters $s$ and $\Omega_{m,0}$ are
constrained to be 
\begin{equation}
s=0.034_{-0.034}^{+0.327}\,,\qquad\Omega_{m,0}=0.271_{-0.010}^{+0.024}\qquad(95\,\%~{\rm CL}).
\end{equation}
 This shows that the tracker solution with $-1.36<w_{{\rm DE}}<-1$
during the matter era can be allowed observationally. The $\Lambda$CDM
model corresponds to the line $s=0$ in Fig.~\ref{Obs-1}, which,
as expected, is inside the 1$\sigma$ contour for $0.264<\Omega_{m,0}<0.273$
(with $\chi_{\Lambda{\rm CDM}}^{2}=469.024$). The best-fit $\chi^{2}$
for the extended Galileon model is slightly smaller than that in the
$\Lambda$CDM model. However, since $\Delta({\rm AIC})=1.85$, the
observational data do not particularly favor the extended Galileon
model over the $\Lambda$CDM model.

\subsubsection{Non-flat case: $\Omega_{K,0}\neq0$}

\begin{figure}
\begin{centering}
\includegraphics[width=4.5in,height=3.5in]{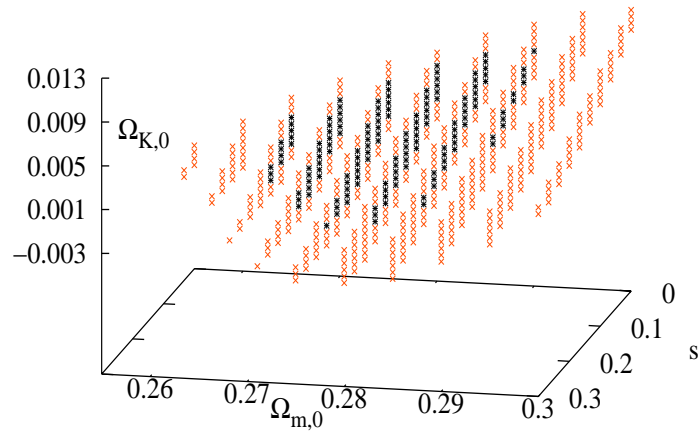} 
\par\end{centering}

\caption{ Observational contours for the tracker solution in the ($s,\Omega_{m,0},\Omega_{K,0}$)
space with the cosmic curvature taken into account. 1$\sigma$ and
2$\sigma$ contours correspond to the black and red regions, respectively.}

\label{Obs-2} 
\end{figure}

In the presence of the cosmic curvature $K$ we also evaluate the
$\chi^{2}$ in the parameter space given by $0\leq s<0.9$, $0.25<\Omega_{m,0}<0.32$,
and $-0.006<\Omega_{K,0}<0.014$. The minimum value of $\chi^{2}$
is found on the point $s=0.067$, $\Omega_{m,0}=0.2768$, and $\Omega_{K,0}=0.00768$,
at which $\chi_{{\rm min},K}^{2}=467.436$. We evaluate the difference
$\Delta\chi_{K}^{2}\equiv\chi^{2}-\chi_{{\rm min},K}^{2}$ on each
grid point, according to which we can exclude: a) the values of $\chi^{2}$
at 68\,\% CL (which, for a system with 3 parameters, corresponds to
$\Delta\chi_{K}^{2}>3.51$), and b) the values of $\chi^{2}$ at 95\%
CL (i.e.\ $\Delta\chi_{K}^{2}>7.82$). In Fig.~\ref{Obs-2} we plot
a three dimensional region (by showing only some slices of it) of
the allowed parameter space for the tracker solution. We find that
the model parameters are constrained to be 
\begin{equation}
s=0.067_{-0.067}^{+0.333}\,,\qquad\Omega_{m,0}=0.277_{-0.022}^{+0.023}\,,\qquad\Omega_{K,0}=0.0077_{-0.0127}^{+0.0039}\qquad
(95\,\%~{\rm CL}).
\end{equation}

The AIC for the best-fit extended Galileon model is ${\rm AIC}=473.436$,
whereas in the best-fit $\Lambda$CDM model with two parameters $\Omega_{m,0}$
and $\Omega_{K,0}$ the AIC is found to be ${\rm AIC}_{\Lambda{\rm CDM},K}=472.543$
with $\chi_{\Lambda{\rm CDM},K}^{2}=468.543$. Since $\Delta({\rm AIC})=0.893$
between the two models, they are equivalently supported by the data.
However, if we compare the non-flat extended Galileon model with the
flat $\Lambda$CDM (1 parameter only, $\Omega_{m,0}$), then we see
that $\Delta({\rm AIC})=2.412$, which states that the flat $\Lambda$CDM
is better supported by the data. According to this criterion, the
flat extended Galileon tracker is favored over the non-flat case from
a statistical point of view.


\section{Cosmological perturbations and the discrimination between the models}

\label{cosmopercon} 

The property of the tracker solution does not depend on the parameters
$\alpha$ and $\beta$ at the level of the background cosmology. If
we consider cosmological perturbations, it is possible to discriminate
between the models with different values of $\alpha$ and $\beta$.
In order to confront the extended Galileon models with the observations
of LSS, CMB, and weak lensing, we shall study the evolution of matter
density perturbations as well as gravitational potentials. Since our
interest is the growth of non-relativistic matter perturbations in
the late Universe, we do not take into account the radiation as far
as the perturbations are concerned.

\subsection{Linear perturbation equations}

Let us consider the perturbed metric in the longitudinal gauge about
the flat FLRW background \cite{Bardeen} %
\begin{equation}
ds^{2}=-(1+2\Psi)\, dt^{2}+a^{2}(t)(1+2\Phi)\delta_{ij}dx^{i}dx^{j}\,,\label{permet}
\end{equation}
where $\Psi$ and $\Phi$ are scalar metric perturbations. We perturb
the scalar field as $\phi(t)+\delta\phi(t,{\bm{x}})$, and non-relativistic
matter as well, in terms of the matter density perturbation $\delta\rho_{m}$
and the scalar part of the fluid velocity $v$. The density contrast
of non-relativistic matter is defined as $\delta\equiv\delta\rho_{m}/\rho_{m}$.
We also introduce the gauge-invariant density contrast %
\begin{equation}
\delta_{m}\equiv\delta+3Hv\,.\label{gaugein}
\end{equation}

The full linear perturbation equations in the Horndeski's most general
scalar-tensor theories were derived in Ref.~\cite{DKT}. For the
extended Galileon models, in Fourier space, they are given by %
\begin{eqnarray}
 &  & A_{1}\dot{\Phi}+A_{2}\dot{\delta\phi}-\rho_{m}\Psi+A_{3}\frac{k^{2}}{a^{2}}\Phi+A_{4}\Psi+A_{6}\frac{k^{2}}{a^{2}}\delta\phi-\rho_{m}\delta=0\,,\label{eq:Psi}\\
 &  & B_{6}\Phi+B_{7}\delta\phi+A_{3}\Psi=0\,,\label{eq:chi}\\
 &  & A_{3}\dot{\Phi}+A_{6}\dot{\delta\phi}-A_{1}\Psi/3+C_{4}\delta\phi+\rho_{m}v=0\,,\label{eq:Phi}\\
 &  & 3A_{6}\ddot{\Phi}+D_{2}\ddot{\delta\phi}+D_{3}\dot{\Phi}+D_{4}\dot{\delta\phi}-A_{2}\dot{\Psi}+\left(B_{7}\frac{k^{2}}{a^{2}}+D_{8}\right)\Phi+D_{9}\frac{k^{2}}{a^{2}}\delta\phi+\left(A_{6}\frac{k^{2}}{a^{2}}+D_{11}\right)\Psi=0\,,\label{eq:dPi}\\
 &  & \dot{v}-\Psi=0\,,\label{eq:mat1}\\
 &  & \dot{\delta}+3\dot{\Phi}+\frac{k^{2}}{a^{2}}v=0\,,\label{eq:mat2}
\end{eqnarray}
 where $k$ is a comoving wave number, and 
\begin{eqnarray}
\hspace{-0.5cm}A_{1} & = & 12HG_{4}-6\,\dot{\phi}XG_{{3,X}}-48\, HX\left(G_{{4,X}}+XG_{{4,{\it XX}}}\right)-6{H}^{2}X\left(5\, G_{{5,X}}+2\, XG_{{5,{\it XX}}}\right)\dot{\phi}\,,\\
\hspace{-0.5cm}A_{2} & = & -\left(K_{,X}+2\, XK_{,XX}\right)\dot{\phi}-6HX\left(3G_{{3,X}}+2XG_{{3,{\it XX}}}\right)-6{H}^{2}\left(3\, G_{{4,X}}+12\, XG_{{4,{\it XX}}}+4{X}^{2}G_{{4,{\it XXX}}}\right)\dot{\phi}\nonumber \\
 &  & {}-2{H}^{3}X\left(15\, G_{{5,X}}+20\, XG_{{5,{\it XX}}}+4\,{X}^{2}G_{{5,{\it XXX}}}\right)\,,\\
\hspace{-0.5cm}A_{3} & = & 4G_{4}-8XG_{{4,X}}-4H\dot{\phi}XG_{{5,X}}\,,\\
\hspace{-0.5cm}A_{4} & = & 2X\left(K_{,X}+2XK_{,XX}\right)-12H^{2}G_{4}+\rho_{m}+12HX\left(2G_{{3,X}}+XG_{{3,{\it XX}}}\right)\dot{\phi}\nonumber \\
 &  & +12{H}^{2}X\left(7G_{{4,X}}+16XG_{{4,{\it XX}}}+4{X}^{2}G_{{4,{\it XXX}}}\right)+4{H}^{3}X\left(15G_{{5,X}}+13XG_{{5,{\it XX}}}+2{X}^{2}G_{{5,{\it XXX}}}\right)\dot{\phi}\,,\\
\hspace{-0.5cm}A_{6} & = & -2XG_{{3,X}}-4H\left(G_{{4,X}}+2XG_{{4,{\it XX}}}\right)\dot{\phi}-2{H}^{2}X\left(3G_{{5,X}}+2XG_{{5,{\it XX}}}\right)\,,\\
\hspace{-0.5cm}B_{6} & = & 4G_{4}-4XG_{5,X}\ddot{\phi}\,,\\
\hspace{-0.5cm}B_{7} & = & -4G_{4,X}H\dot{\phi}-4(G_{4,X}+2XG_{4,XX})\ddot{\phi}-4(G_{5,X}+XG_{5,XX})H\dot{\phi}\ddot{\phi}-4XG_{5,X}(H^{2}+\dot{H})\,,\\
\hspace{-0.5cm}C_{4} & = & K_{,X}\dot{\phi}+6HXG_{3,X}+6H^{2}(G_{4,X}+2XG_{4,XX})\dot{\phi}+2XH^{3}(3G_{5,X}+2XG_{5,XX})\,,\\
\hspace{-0.5cm}D_{9} & = & -K_{,X}-2(G_{3,X}+XG_{3,XX})\ddot{\phi}-4HG_{3,X}\dot{\phi}-4H(3G_{4,XX}+2XG_{4,XXX})\dot{\phi}\ddot{\phi}-2G_{4,X}(3H^{2}+2\dot{H})\nonumber \\
 &  & -4XG_{4,XX}(5H^{2}+2\dot{H})-2H^{2}(G_{5,X}+5XG_{5,XX}+2X^{2}G_{5,XXX})\ddot{\phi}-4H(H^{2}+\dot{H})(G_{5,X}+XG_{5,XX})\dot{\phi}.
\end{eqnarray}
 The readers may refer to the paper \cite{DKT} for the explicit
forms of the coefficients $D_{2},D_{3},D_{4},D_{8}$, and $D_{11}$.
Since the perturbation equations (\ref{eq:Psi})-(\ref{eq:mat2})
are not independent, we do not need to know these unwritten coefficients
to solve the equations numerically. Now we are dealing with a massless
scalar field, so that the mass term $M$ does not appear in the perturbation
equations.

{}From Eqs.~(\ref{eq:mat1}) and (\ref{eq:mat2}) the gauge-invariant
matter perturbation (\ref{gaugein}) obeys %
\begin{equation}
\ddot{\delta}_{m}+2H\dot{\delta}_{m}+\frac{k^{2}}{a^{2}}\Psi=3\left(\ddot{I}+2H\dot{I}\right)\,,\label{delmeq}
\end{equation}
where $I\equiv Hv-\Phi$. We define the effective gravitational potential
\begin{equation}
\Phi_{{\rm eff}}\equiv(\Psi-\Phi)/2\,.\label{Phieff}
\end{equation}
 This quantity is related to the deviation of the light rays in CMB
and weak lensing observations \cite{Schmi}. To quantify the difference
between the two gravitational potentials $\Phi$ and $\Psi$, we also
introduce 
\begin{equation}
\eta\equiv-\Phi/\Psi\,,\label{eta}
\end{equation}
by which Eq.~(\ref{Phieff}) can be written as $\Phi_{{\rm eff}}=\Psi(1+\eta)/2$.

\subsection{Quasi-static approximation on sub-horizon scales}

For the modes deep inside the Hubble radius ($k^{2}/a^{2}\gg H^{2}$)
we can employ the quasi-static approximation under which the dominant
contributions in the perturbation equations are those including $k^{2}/a^{2}$
and $\delta$ (or $\delta_{m}$) \cite{quasi,Kase}. This approximation
is known to be trustable as long as the oscillating mode of the field
perturbation is negligible relative to the matter-induced mode. Combining
Eqs.~(\ref{eq:Psi}), (\ref{eq:chi}), and (\ref{eq:dPi}) under
this approximation, it follows that \cite{DKT} %
\begin{equation}
\frac{k^{2}}{a^{2}}\Psi\simeq-4\pi G_{{\rm eff}}\rho_{m}\delta\,,\label{Poi}
\end{equation}
where %
\begin{equation}
G_{{\rm eff}}=\frac{2M_{{\rm pl}}^{2}(B_{6}D_{9}-B_{7}^{2})}{A_{6}^{2}B_{6}+A_{3}^{2}D_{9}-2A_{3}A_{6}B_{7}}G\,.\label{Geff}
\end{equation}
Here we introduced the bare gravitational constant $G=1/(8\pi M_{{\rm pl}}^{2})$.
Substituting Eq.~(\ref{Poi}) into Eq.~(\ref{delmeq}) with the
relation $\delta_{m}\simeq\delta$ (valid for $k^{2}/a^{2}\gg H^{2}$),
we obtain 
\begin{equation}
\delta_{m}''+\left(2+\frac{H'}{H}\right)\delta_{m}'-\frac{3}{2}\frac{G_{{\rm eff}}}{G}\Omega_{m}\delta_{m}\simeq0\,,\label{delmeq2}
\end{equation}
 where a prime represents a derivative with respect to $N=\ln a$.

Under the quasi-static approximation the quantity $\eta$ defined
in Eq.~(\ref{eta}) reads \cite{DKT} 
\begin{equation}
\eta\simeq\frac{A_{3}D_{9}-A_{6}B_{7}}{B_{6}D_{9}-B_{7}^{2}}\,.
\end{equation}
 On using Eq.~(\ref{Poi}), the effective gravitational potential
$\Phi_{{\rm eff}}=\Psi(1+\eta)/2$ yields 
\begin{equation}
\Phi_{{\rm eff}}\simeq-\frac{3}{2}\xi\left(\frac{aH}{k}\right)^{2}\Omega_{m}\delta_{m}\,,
\end{equation}
 where 
\begin{equation}
\xi\equiv\frac{G_{{\rm eff}}}{G}\frac{1+\eta}{2}\simeq\frac{M_{{\rm pl}}^{2}(B_{6}D_{9}-B_{7}^{2}+A_{3}D_{9}-A_{6}B_{7})}{A_{6}^{2}B_{6}+A_{3}^{2}D_{9}-2A_{3}A_{6}B_{7}}\,.\label{xi}
\end{equation}
 The $\Lambda$CDM model corresponds to $K=-\Lambda$, $G_{4}=M_{{\rm pl}}^{2}/2$,
$G_{3}=G_{5}=0$, in which case $A_{3}=B_{6}=2M_{{\rm pl}}^{2}$,
$A_{6}=0$, and $B_{7}=0$. Then one has $G_{{\rm eff}}/G=1$ and
$\xi=1$ from Eqs.~(\ref{Geff}) and (\ref{xi}).

In the extended Galileon models the general expressions of $G_{{\rm eff}}/G$
and $\xi$ are quite complicated. In what follows we shall focus on
the evolution of cosmological perturbations for the tracker solution
($r_{1}=1$).

In the early cosmological epoch ($r_{2}\ll1$) and during the matter
domination ($\Omega_{r}=0$), $G_{{\rm eff}}/G$ and $\xi$ are approximately
given by 
\begin{eqnarray}
G_{{\rm eff}}/G & \simeq & 1+[27p(2p-1)(3\alpha^{2}p+6\beta^{2}(2p-1)+\beta(1+3\alpha-2(1+6\alpha)p))+9(6\alpha^{2}p(18p-5)\nonumber \\
 &  & +2\beta(2p-1)(\beta(90p-3)-11p)+\alpha(2p(2p-1)-3\beta(1+4p(33p-13))))q+2(-9(9\alpha-23\beta)(\alpha-2\beta)\nonumber \\
 &  & +6(147\alpha^{2}+\alpha(5-507\beta)+6\beta(71\beta-3))p+8(9\alpha-18\beta-1)p^{2}+16p^{3})q^{2}+4(9(27\alpha-46\beta)(\alpha-2\beta)\nonumber \\
 &  & -12(9\alpha-18\beta-8)(\alpha-2\beta)p+16(1-3\alpha+6\beta)p^{2})q^{3}+48(\alpha-2\beta)(6\beta(7-2p)-8p+3\alpha(2p-7))q^{4}\nonumber \\
 &  & +576(\alpha-2\beta)^{2}q^{5}]r_{2}/\Delta,\label{Geffap}\\
\xi & \simeq & 1+[27p(2p-1)(3\alpha^{2}p+6\beta^{2}(2p-1)+\beta(1+3\alpha-2(1+6\alpha)p))+9(24\alpha^{2}p(4p-1)\nonumber \\
 &  & +2\beta(2p-1)(\beta(78p+3)-11p)+\alpha(4p(2p-1)+\beta(3+12(10-29p)p)))q+2(-45(\alpha-2\beta)\beta\nonumber \\
 &  & +6(96\alpha^{2}+\alpha(10-321\beta)+6\beta(43\beta-2))p+4(27\alpha-54\beta-4)p^{2}+32p^{3})q^{2}+8(9(3\alpha-10\beta)(\alpha-2\beta)\nonumber \\
 &  & -3(27\alpha-54\beta-26)(\alpha-2\beta)p+16(1-3\alpha+6\beta)p^{2})q^{3}+48(\alpha-2\beta)(66\beta-8(2+3\beta)p+3\alpha(4p-11))q^{4}\nonumber \\
 &  & +1152(\alpha-2\beta)^{2}q^{5}]r_{2}/(2\Delta),\label{xiap}
\end{eqnarray}
 where 
\begin{eqnarray}
\Delta & \equiv & 4q[(3-9\alpha)p+12p^{3}+2pq(\alpha(60-66q)+20q-11)+2p^{2}(\alpha(9-18q)+22q-6)-3\alpha q(9-54q+40q^{2})\nonumber \\
 &  & +3\beta(2p+4q-1)(3-22q+20q^{2}+6p(2q-1))]\,.
\end{eqnarray}
 To derive Eqs.~(\ref{Geffap}) and (\ref{xiap}) we set $\Omega_{r}=0$
and performed the Taylor expansion around $r_{2}=0$.

At the de Sitter solution ($r_{1}=r_{2}=1$), $G_{{\rm eff}}$ and
$\xi$ are simply given by 
\begin{equation}
G_{{\rm eff}}/G=\xi=\frac{2}{2(1-p)+3(1+2q)(\alpha-2\beta)}\,.\label{Geffde}
\end{equation}
 The equality of $G_{{\rm eff}}/G$ and $\xi$ comes from the fact
that $\eta=1$ at $r_{1}=r_{2}=1$.

Let us first consider the theory where $\alpha=\beta=0$. {}From
Eqs.~(\ref{Geffap})-(\ref{Geffde}) it follows that 
\begin{eqnarray}
 &  & G_{{\rm eff}}/G=\xi\simeq1+\frac{4pq}{6p+10q-3}r_{2}\qquad\quad({\rm for}~~r_{2}\ll1),\\
 &  & G_{{\rm eff}}/G=\xi=\frac{1}{1-p}\qquad\qquad\qquad\qquad\quad({\rm for}~~r_{2}=1).
\end{eqnarray}
 For $p=1$ \cite{Kimura} both $G_{{\rm eff}}/G$ and $\xi$ diverge
at the de Sitter solution. In this case one has $G_{{\rm eff}}/G=\xi\simeq1+4qr_{2}/(10q+3)$
in the regime $r_{2}\ll1$, so that $G_{{\rm eff}}/G$ and $\xi$
are larger than 1 for $q>0$. The property that $\xi$ is as large
as $G_{{\rm eff}}/G$ leads to the enhancement of the effective gravitational
potential relative to the matter perturbation $\delta_{m}$ normalized
by $a$ \cite{Kimura}. This gives rise to the anti-correlation between
the late-time ISW effect and the LSS. Then the parameter $q$ is constrained
to be $q>4.2\times10^{3}$ at the 95 \% confidence level \cite{Kimura2},
which means that $w_{{\rm DE}}$ is very close to $-1$ along the
tracker.

The situation is different for $\alpha\neq0$ and $\beta\neq0$. Let
us consider the models with $p=1$ and $q=5/2$, i.e. $s=0.2$, in
which case the models are compatible with the observational constraints
discussed in Sec.~\ref{obsercon}. In the regime $r_{2}\ll1$, Eqs.~(\ref{Geffap})
and (\ref{xiap}) read 
\begin{eqnarray}
G_{{\rm eff}}/G & \simeq & 1+\frac{48411\alpha^{2}-3\alpha(3560+60291\beta)+22(50+924\beta+7641\beta^{2})}{10(308-1536\alpha+3201\beta)}r_{2}\,,\label{Geff1}\\
\xi & \simeq & 1+\frac{75276\alpha^{2}-3\alpha(8020+99981\beta)+22[100+3\beta(733+4527\beta)]}{20(308-1536\alpha+3201\beta)}r_{2}\,.\label{xi1}
\end{eqnarray}
 At the de Sitter solution Eq.~(\ref{Geffde}) gives 
\begin{equation}
G_{{\rm eff}}/G=\xi=\frac{1}{9(\alpha-2\beta)}\,.\label{Geffde2}
\end{equation}

\begin{figure}
\begin{centering}
\includegraphics[width=3.5in,height=2.6in]{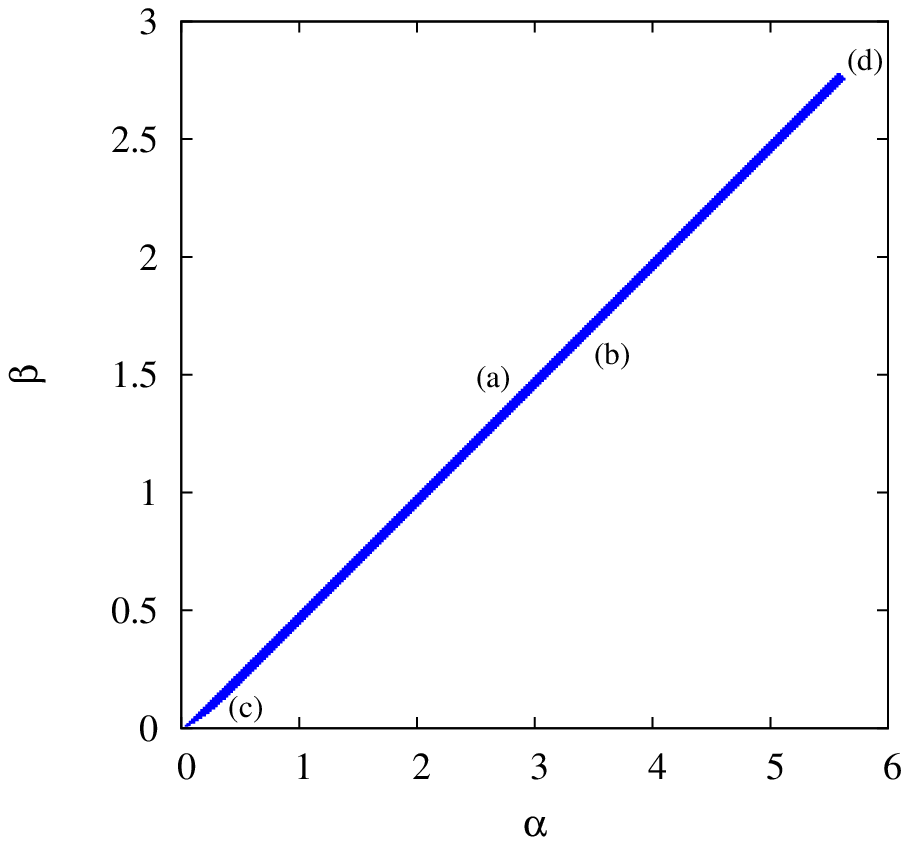} \includegraphics[width=3.5in,height=2.6in]{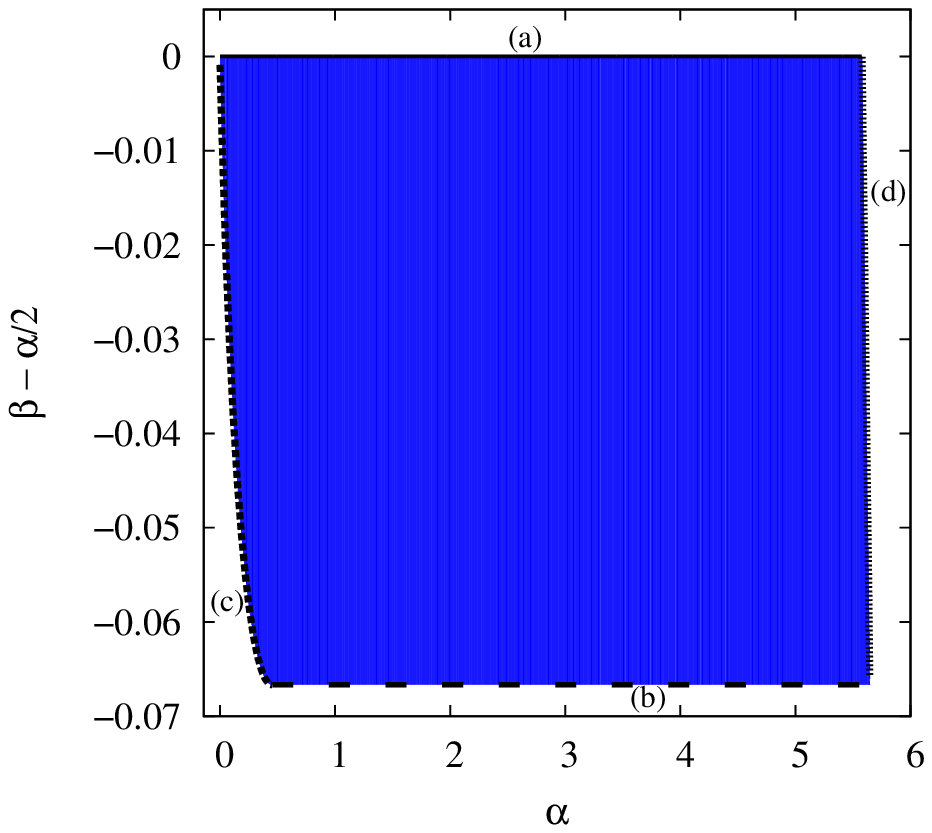} 
\par\end{centering}

\caption{Allowed parameter space in the ($\alpha,\beta$) plane for $p=1$
and $q=5/2$. In the blue region the conditions for the avoidance
of ghosts and Laplacian instabilities of scalar and tensor perturbations
are satisfied. The right panel shows the enlarged version of the left
panel in the $(\alpha,\beta-\alpha/2)$ plane. The borders correspond
to (a) $\beta=\alpha/2$, (b) $\beta=\alpha/2-1/15$, (c) $\beta=(408\alpha+68-2\sqrt{17}\sqrt{3(272-75\alpha)\alpha+68})/561$,
and (d) $\beta=(242-15\alpha+4\sqrt{3630-495\alpha})/99$, respectively.
Taken from Ref.~\cite{Defe2011}.}

\centering{}\label{validfig} 
\end{figure}


In Ref.~\cite{Defe2011} the authors clarified the viable parameter
region in which the ghosts and Laplacian instabilities of scalar and
tensor perturbations are absent. In Fig.~\ref{validfig} we plot
the allowed parameter space in the ($\alpha,\beta$) plane for $p=1$
and $q=5/2$. The viable region is surrounded by the four borders:
(a) $\beta=\alpha/2$, (b) $\beta=\alpha/2-1/15$, (c) $\beta=(408\alpha+68-2\sqrt{17}\sqrt{3(272-75\alpha)\alpha+68})/561$,
and (d) $\beta=(242-15\alpha+4\sqrt{3630-495\alpha})/99$. The model
with $\alpha=\beta=0$ is on the border lines {[}the intersection
of the lines (a) and (c){]}, in which case both $G_{{\rm eff}}/G$
and $\xi$ diverge at the de Sitter solution.

\begin{figure}
\begin{centering}
\includegraphics[width=3.5in,height=2.7in]{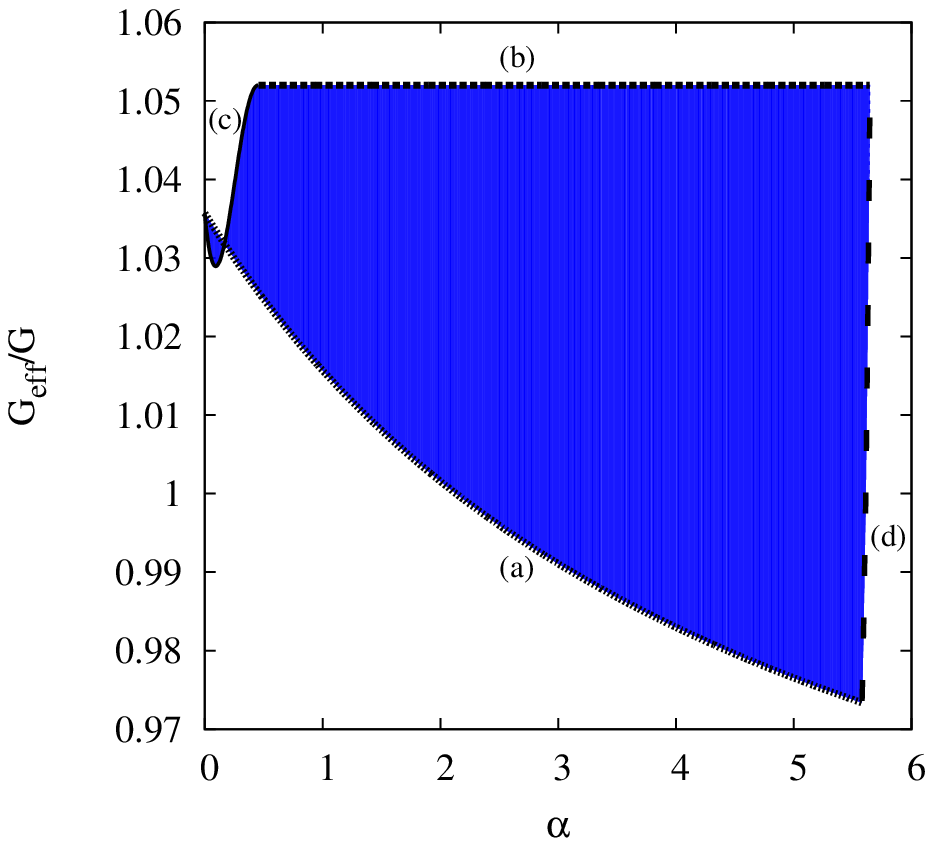} \includegraphics[width=3.5in,height=2.7in]{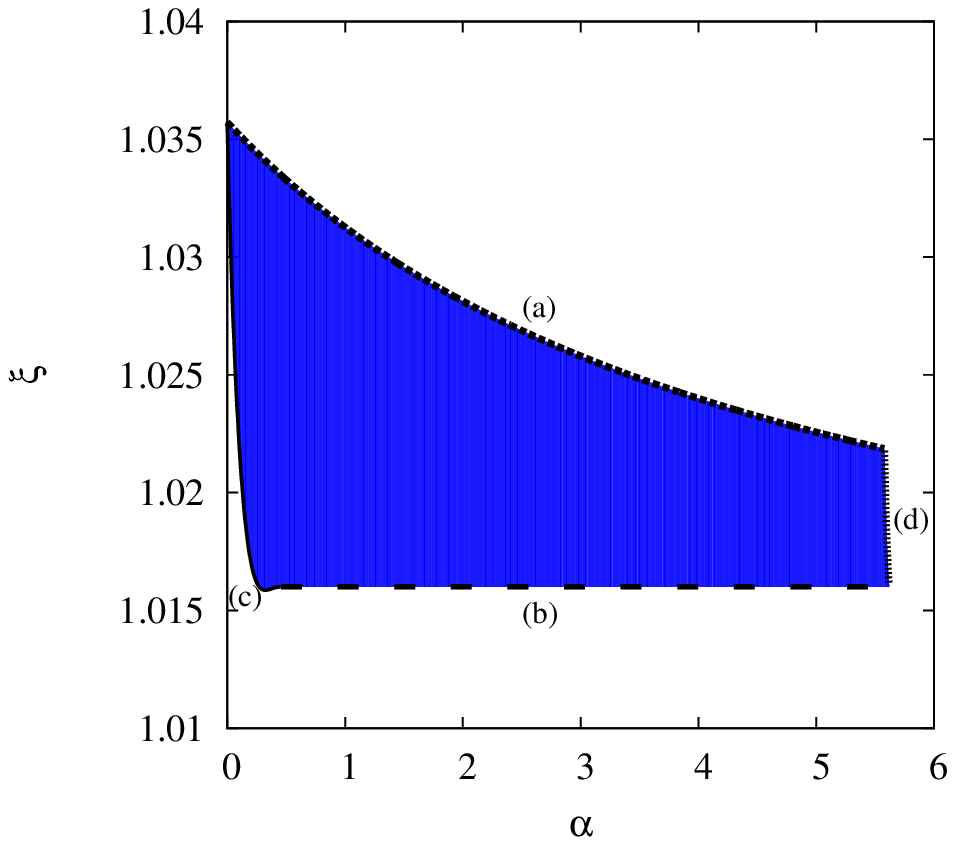} 
\par\end{centering}

\caption{The left and right panels show the values of $G_{{\rm eff}}/G$ and
$\xi=(G_{{\rm eff}}/G)(1+\eta)/2$ versus $\alpha$ for $p=1$ and
$q=5/2$, respectively, along the tracker solution ($r_{1}=1$) at
$r_{2}=0.1$. The blue regions illustrate the viable parameter spaces
in which the parameters $\alpha$ and $\beta$ belong to the blue
region in Fig.~\ref{validfig}. The borders (a), (b), (c), (d) correspond
to those given in Fig.~\ref{validfig}.}

\centering{}\label{earlyfig} 
\end{figure}


Figure \ref{earlyfig} illustrates the regions for the possible values
of $G_{{\rm eff}}/G$ in Eq.~(\ref{Geff1}) and $\xi$ in Eq.~(\ref{xi1})
for $p=1$ and $q=5/2$ at $r_{2}=0.1$. The blue regions are surrounded
by the borders (a), (b), (c), (d), which correspond to those given
in Fig.~\ref{validfig} respectively. The lines (a) in Fig.~\ref{earlyfig}
show $G_{{\rm eff}}/G$ and $\xi$ for $\beta=\alpha/2$, along which
both $G_{{\rm eff}}/G$ and $\xi$ decrease for larger $\alpha$.
One has $G_{{\rm eff}}/G=\xi=1.035$ for $\alpha=\beta=0$, in which
case our numerical simulations show that there is an anti-correlation
between $\delta_{m}/a$ and $\Phi_{{\rm eff}}$ for the modes deep
inside the horizon. When $\alpha>0$ the variable $\xi$ is larger
than $G_{{\rm eff}}/G$ on the line (a) for the same $\alpha$, so
that the anti-correlation tends to be even stronger than that for
$\alpha=0$.

On the line (b) in Fig.~\ref{validfig}, i.e. $\beta=\alpha/2-1/15$,
one has $G_{{\rm eff}}/G\simeq1+0.52r_{2}$ and $\xi=1+0.16r_{2}$,
which are independent of the values of $\alpha$. The fact that $G_{{\rm eff}}/G$
is larger than $\xi$ may imply the absence of the anti-correlation
between $\delta_{m}/a$ and $\Phi_{{\rm eff}}$. In fact we will show
in Sec.~\ref{numerics} that the anti-correlation tends to disappear
as the model parameters approach the border line (b) in Fig.~\ref{validfig}.
For given $\alpha$, as we move from $\beta=\alpha/2$ to $\beta=\alpha/2-1/15$,
$G_{{\rm eff}}/G$ increases whereas $\xi$ gets smaller. For $0\leq\alpha\leq34/75$
the viable regions for $G_{{\rm eff}}/G$ and $\xi$ in Fig.~\ref{earlyfig}
are surrounded by the lines (a), (c), and $\alpha=34/75$, whereas
for $5.579\leq\alpha\leq5.646$ they are surrounded by the lines (b),
(d), and $\alpha=5.579$.

\begin{figure}
\begin{centering}
\includegraphics[width=3.8in,height=2.8in]{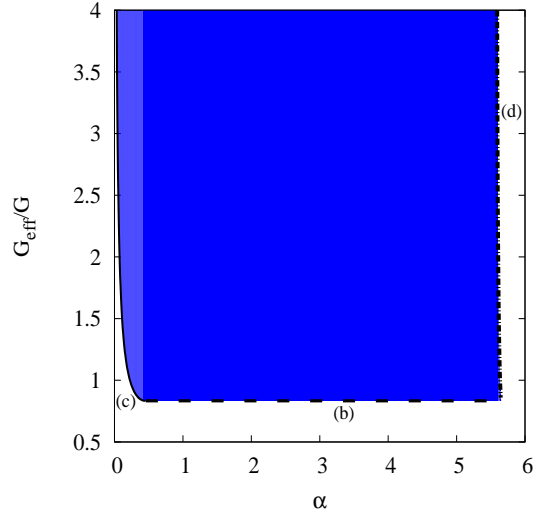} 
\par\end{centering}

\caption{$G_{{\rm eff}}/G$ versus $\alpha$ for $p=1$ and $q=5/2$ at the
de Sitter solution ($r_{1}=1,r_{2}=1$). Note that the parameter $\xi$
is exactly the same as $G_{{\rm eff}}/G$. $G_{{\rm eff}}/G$ diverges
at $\beta=\alpha/2$, so that it is unbounded from above. The borders
(b), (c), (d) correspond to those given in Fig.~\ref{validfig},
respectively.}

\centering{}\label{desfig} 
\end{figure}


In Fig.~\ref{desfig} we plot $G_{{\rm eff}}/G=\xi=1/[9(\alpha-2\beta)]$
for the late-time de Sitter solution. Both $G_{{\rm eff}}/G$ and
$\xi$ diverge on the line (a) in Fig.~\ref{validfig}, including
the case $\alpha=\beta=0$. On the line (b) one has $G_{{\rm eff}}/G=\xi=5/6$,
so that the growth rate of matter perturbations is smaller than that
in the $\Lambda$CDM model around the future de Sitter solution. The
allowed values of $G_{{\rm eff}}/G$ and $\xi$ exist in the wide
regions in Fig.~\ref{desfig}.

The present epoch corresponds to the regime between $r_{2}\ll1$ and
$r_{2}=1$, so we need to resort to numerical simulations to estimate
the growth rate of perturbations accurately.

\subsection{Numerical simulations}

\label{numerics}

In order to study the evolution of perturbations for a number of different
wave numbers (both sub-horizon and super-horizon modes), we shall
solve the full perturbation equations numerically without using the
quasi-static approximation. Let us introduce the following dimensionless
variables 
\begin{equation}
V\equiv Hv\,,\qquad\delta\varphi\equiv\delta\phi/(x_{{\rm dS}}M_{{\rm pl}})\,,
\end{equation}
 with $\tilde{A}_{1}\equiv A_{1}/(HM_{{\rm pl}}^{2})$, $\tilde{A}_{2}\equiv x_{{\rm dS}}A_{2}/(HM_{{\rm pl}})$,
$\tilde{A}_{3}\equiv A_{3}/M_{{\rm pl}}^{2}$, $\tilde{A}_{4}\equiv A_{4}/(H^{2}M_{{\rm pl}}^{2})$,
$\tilde{A}_{6}\equiv x_{{\rm dS}}A_{6}/M_{{\rm pl}}$, $\tilde{B}_{6}\equiv B_{6}/M_{{\rm pl}}^{2}$,
$\tilde{B}_{7}\equiv x_{{\rm dS}}B_{7}/M_{{\rm pl}}$, and $\tilde{C}_{4}\equiv x_{{\rm dS}}C_{4}/(HM_{{\rm pl}})$.
{}From Eqs.~(\ref{eq:Psi}), (\ref{eq:chi}), (\ref{eq:Phi}), (\ref{eq:mat1}),
and (\ref{eq:mat2}) we obtain 
\begin{eqnarray}
 &  & \Psi=-(\tB_{6}\Phi+\tB_{7}\delta\varphi)/\tA_{3}\,,\label{per1}\\
 &  & \Phi'=[(3\tA_{4}\tA_{6}\tB_{6}+\tA_{1}\tA_{2}\tB_{6}-3\tA_{3}^{2}\tA_{6}k^{2}/(aH)^{2}-9\tA_{6}\tB_{6}\Omega_{m})\Phi\nonumber \\
 &  & ~~~~~~~+(3\tA_{2}\tA_{3}\tC_{4}+3\tA_{4}\tA_{6}\tB_{7}+\tA_{1}\tA_{2}\tB_{7}-9\tA_{6}\tB_{7}\Omega_{m}-3\tA_{3}\tA_{6}^{2}k^{2}/(aH)^{2})\delta\varphi+9\tA_{3}\tA_{6}\Omega_{m}\delta+9\tA_{2}\tA_{3}\Omega_{m}V]\nonumber \\
 &  & ~~~~~~~\times[3\tA_{3}(\tA_{1}\tA_{6}-\tA_{2}\tA_{3})]^{-1}\,,\label{per2}\\
 &  & \delta\varphi'=-[(\tA_{1}^{2}\tB_{6}+3\tA_{3}\tA_{4}\tB_{6}-3\tA_{3}^{3}k^{2}/(aH)^{2}-9\tA_{3}\tB_{6}\Omega_{m})\Phi\nonumber \\
 &  & ~~~~~~~~+(\tA_{1}^{2}\tB_{7}+3\tA_{3}\tA_{4}\tB_{7}+3\tA_{1}\tA_{3}\tC_{4}-3\tA_{3}^{2}\tA_{6}k^{2}/(aH)^{2}-9\tA_{3}\tB_{7}\Omega_{m})\delta\varphi+9\tA_{3}^{2}\Omega_{m}\delta+9\tA_{1}\tA_{3}\Omega_{m}V]\nonumber \\
 &  & ~~~~~~~~\times[3\tA_{3}(\tA_{1}\tA_{6}-\tA_{2}\tA_{3})]^{-1}\,,\label{per3}\\
 &  & \delta'=-3\Phi'-k^{2}/(aH)^{2}\, V\,,\label{per4}\\
 &  & V'=(H'/H)V+\Psi\,.\label{per5}
\end{eqnarray}

In order to recover the General Relativistic behavior in the early
cosmological epoch we choose the initial conditions $\Phi'=0$, $\delta\varphi'=0$,
$\delta\varphi=0$, and $\delta=10^{-5}$, in which case $\Phi_{i}$,
$\Psi_{i}$, and $V_{i}$ are known from Eq.~(\ref{per1})-(\ref{per3})
(where the subscript ``$i$'' represents the initial values). For
non-zero initial values of $\delta\varphi$ the field perturbation
oscillates at the early stage. For the initial conditions with $|\delta\varphi_{i}|\lesssim|\Psi_{i}|$
the evolution of perturbations in the low-redshift regime is hardly
affected by the oscillations. This situation is similar to that found
in Ref.~\cite{Kimura} for the model with $\alpha=\beta=0$.

\begin{figure}
\begin{centering}
\includegraphics[width=3.2in,height=3.1in]{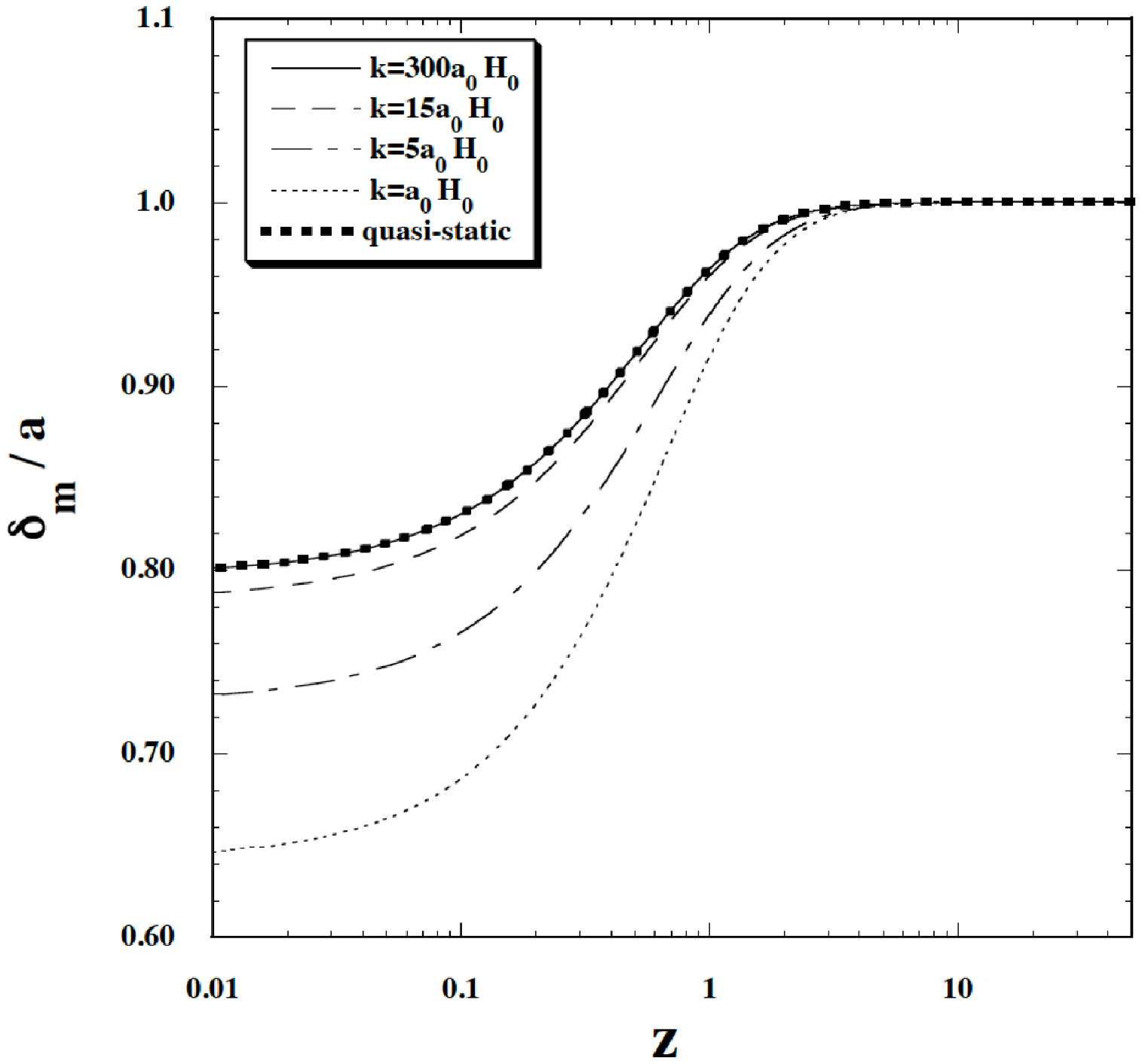} \includegraphics[width=3.2in,height=3.1in]{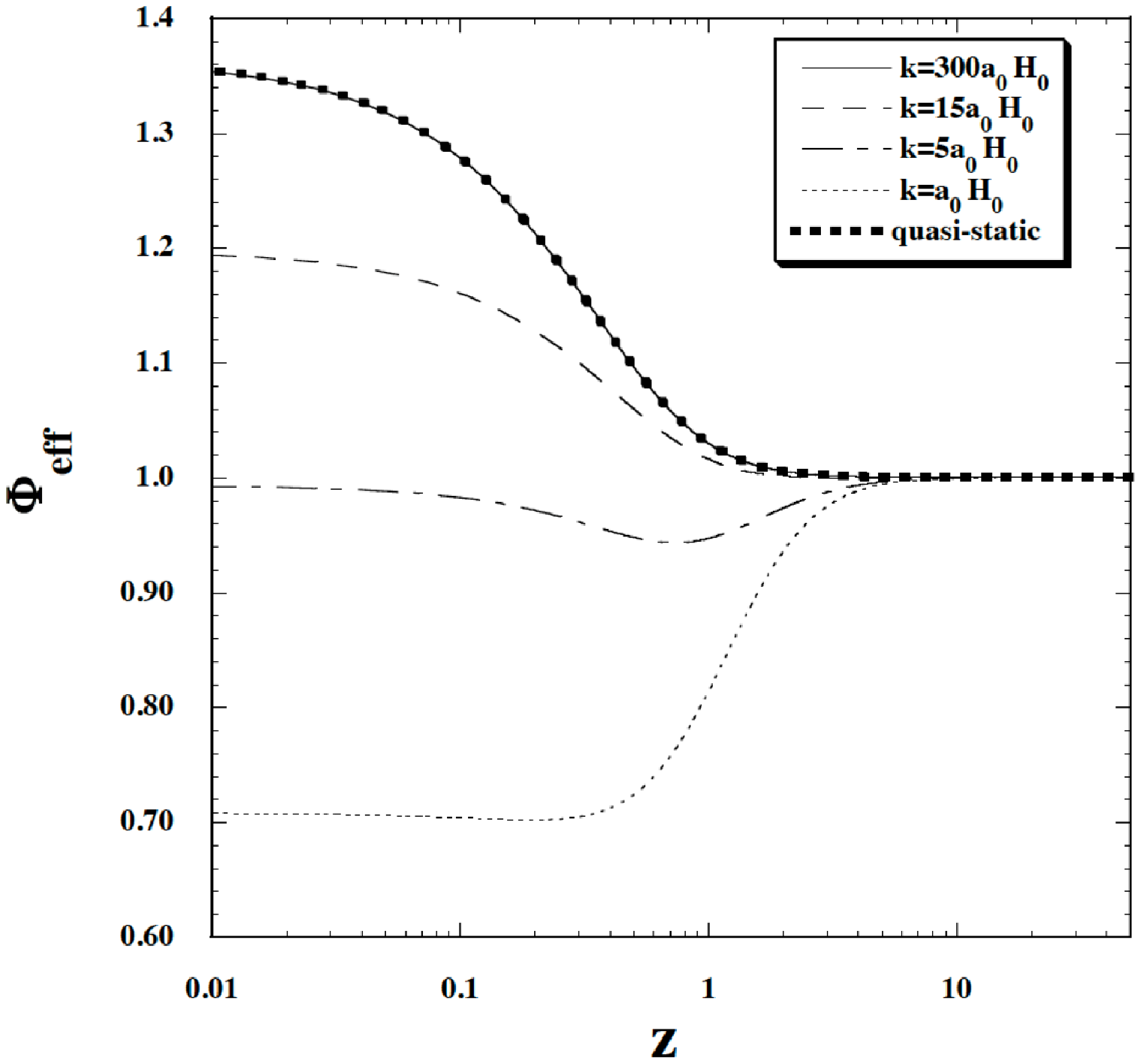} 
\par\end{centering}

\caption{Evolution of $\delta_{m}/a$ (left) and $\Phi_{{\rm eff}}$ (right)
versus the redshift $z$ for $p=1$, $q=5/2$, $\alpha=3$, $\beta=1.49$
along the tracker solution. Note that $\delta_{m}/a$ and $\Phi_{{\rm eff}}$
are normalized by their initial values, respectively. Each curve corresponds
to the evolution of perturbations for the wave numbers $k=300a_{0}H_{0}$,
$k=15a_{0}H_{0}$, $k=5a_{0}H_{0}$, and $k=a_{0}H_{0}$. The dotted
curves show the results obtained under the quasi-static approximation
on sub-horizon scales. In this case there is an anti-correlation between
$\delta_{m}/a$ and $\Phi_{{\rm eff}}$ for the modes $k\gtrsim5a_{0}H_{0}$.}

\centering{}\label{anfig} 
\end{figure}


\begin{figure}
\begin{centering}
\includegraphics[width=3.2in,height=3.1in]{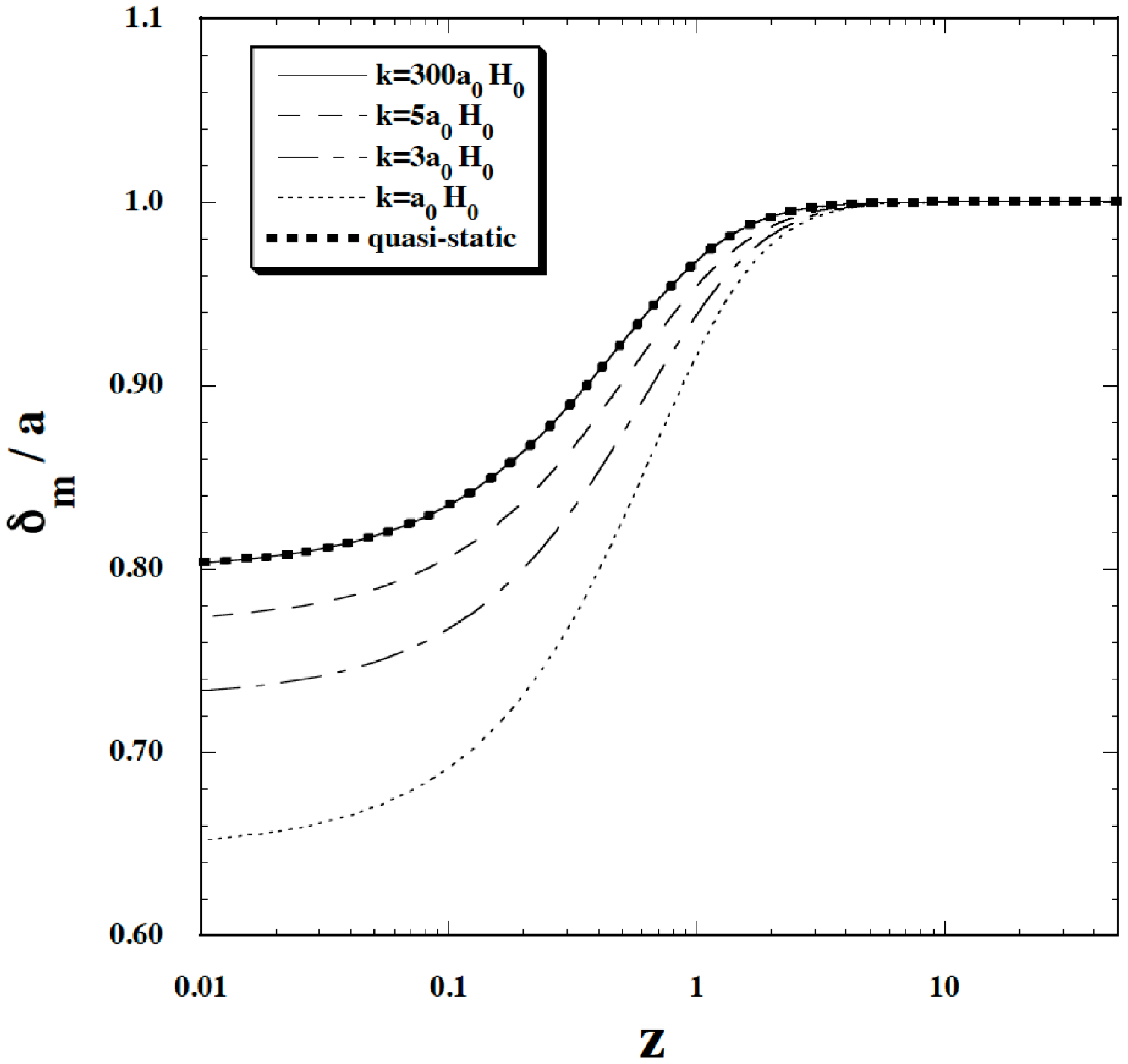} \includegraphics[width=3.2in,height=3.1in]{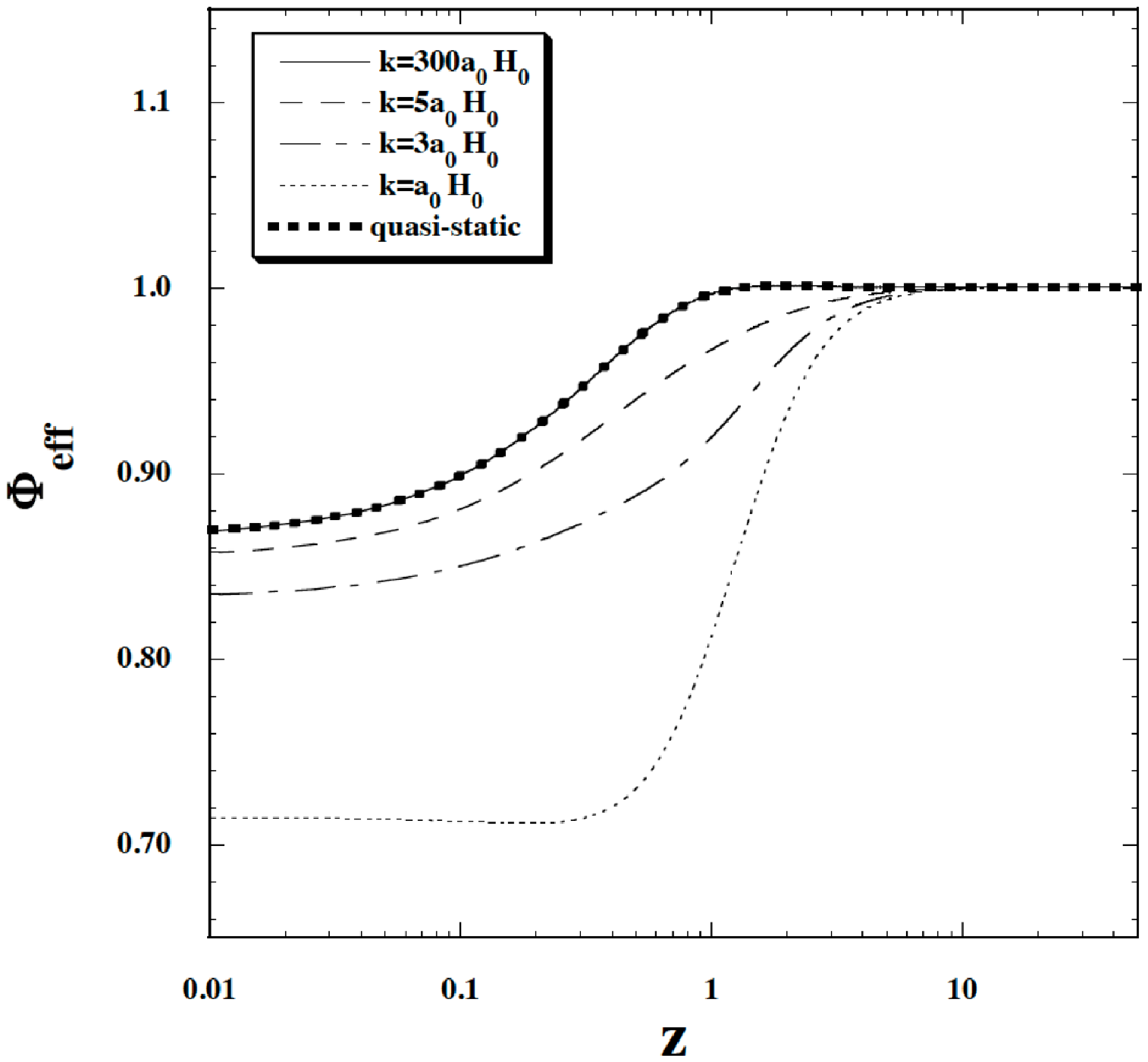} 
\par\end{centering}

\caption{Similar to Fig.~\ref{anfig}, but for the model parameters $p=1$,
$q=5/2$, $\alpha=3$, $\beta=1.45$. In this case $\delta_{m}/a$
and $\Phi_{{\rm eff}}$ are not anti-correlated.}

\centering{}\label{pofig} 
\end{figure}


In Fig.~\ref{anfig} we plot the evolution of $\delta_{m}/a$ and
$\Phi_{{\rm eff}}$ versus the redshift $z=1/a-1$ for $p=1$, $q=5/2$,
$\alpha=3$, and $\beta=1.49$ with several different wave numbers.
In this case the model parameters are close to the border line (a)
in Fig.~\ref{validfig}. The solid curve in Fig.~\ref{anfig} corresponds
to the simulation for the wave number $k=300\, a_{0}H_{0}\simeq0.1\, h$\,Mpc$^{-1}$,
where the subscript ``0'' represents the today's values. The numerical
results of $\delta_{m}/a$ and $\Phi_{{\rm eff}}$ show good agreement
with those derived under the quasi-static approximation on sub-horizon
scales. For the model parameters used in Fig.~\ref{anfig} the analytic
estimates (\ref{Geff1}) and (\ref{xi1}) give $G_{{\rm eff}}/G\simeq1-0.05r_{2}$
and $\xi\simeq1+0.23r_{2}$ in the regime $r_{2}\ll1$, whereas from
Eq.~(\ref{Geffde2}) one has $G_{{\rm eff}}/G=\xi=5.56$ at the de
Sitter solution. {}From Fig.~\ref{anfig} we find that $\delta_{m}/a$
is anti-correlated with $\Phi_{{\rm eff}}$ for the modes $k\gg a_{0}H_{0}$.
This property comes from the fact that the parameter $\xi$ is always
larger than 1, while $G_{{\rm eff}}/G$ is smaller than $\xi$ in
the regime $r_{2}\ll1$. In Fig.~\ref{anfig} the growth rates of
$\delta_{m}$ and $\Phi_{{\rm eff}}$ decrease for smaller $k$. The
anti-correlation between $\delta_{m}/a$ and $\Phi_{{\rm eff}}$ is
present for the modes $k\gtrsim5a_{0}H_{0}$.

Figure \ref{pofig} shows the evolution of $\delta_{m}/a$ and $\Phi_{{\rm eff}}$
for $p=1$, $q=5/2$, $\alpha=3$, and $\beta=1.45$ with several
different wave numbers. In this case $\beta$ is smaller than that
used in Fig.~\ref{anfig}. {}From Eqs.~(\ref{Geff1}) and (\ref{xi1})
one has $G_{{\rm eff}}/G=1+0.25r_{2}$ and $\xi=1+0.16r_{2}$ for
$r_{2}\ll1$, whereas $G_{{\rm eff}}=\xi=1.11$ at the de Sitter solution.
Compared to the case $\beta=1.49$, $G_{{\rm eff}}/G$ and $\xi$
get larger and smaller, respectively, for the same value of $r_{2}$
($\ll1$). This leads to the suppression of the growth of $\Phi_{{\rm eff}}$.
{}From Fig.~\ref{pofig} we find that $\delta_{m}/a$ and $\Phi_{{\rm eff}}$
are positively correlated for the mode $k=300a_{0}H_{0}$. For the
wave numbers $k\gg a_{0}H_{0}$ the quasi-static approximation reproduces
the numerical results in high accuracy. On larger scales both $\delta_{m}$
and $\Phi_{{\rm eff}}$ evolve more slowly, so that $\delta_{m}/a$
and $\Phi_{{\rm eff}}$ are also positively correlated.

\begin{figure}
\begin{centering}
\includegraphics[width=3.2in,height=3.1in]{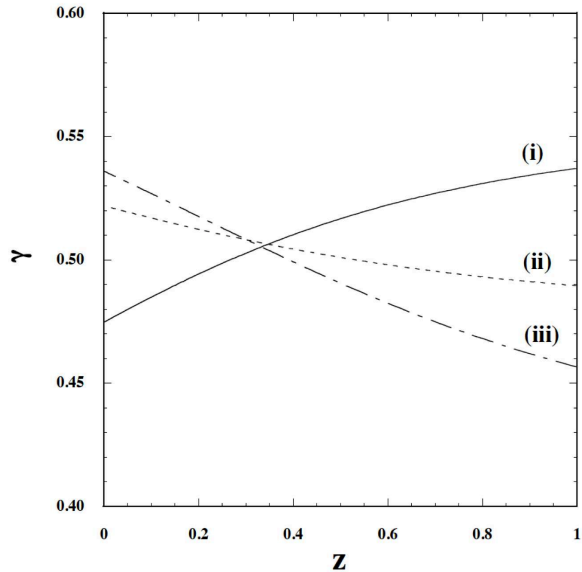} 
\par\end{centering}

\caption{Evolution of the growth index $\gamma$ for the mode $k=300a_{0}H_{0}$
in the regime $0<z<1$ with the model parameters (i) $p=1$, $q=5/2$,
$\alpha=3$, $\beta=1.49$, (ii) $p=1$, $q=5/2$, $\alpha=3$, $\beta=1.45$,
and (iii) $p=1$, $q=5/2$, $\alpha=3$, $\beta=1.434$.}

\centering{}\label{gammafig} 
\end{figure}


We define growth index $\gamma$ of matter perturbations, as \cite{Peebles}
\begin{equation}
\frac{\dot{\delta}_{m}}{H\delta_{m}}=(\Omega_{m})^{\gamma}\,.
\end{equation}
 In the $\Lambda$CDM model $\gamma\simeq0.55$ for the redshift $0\le z\lesssim1$
\cite{Wang98}. In Fig.~\ref{gammafig} we plot the evolution of
$\gamma$ for $p=1$ and $q=5/2$ with the wave number $k=300a_{0}H_{0}$
in three different cases: (i) $\alpha=3$, $\beta=1.49$, (ii) $\alpha=3$,
$\beta=1.45$, and (iii) $\alpha=3$, $\beta=1.434$. The numerical
simulations for (i) and (ii) correspond to the model parameters used
in Figs.~\ref{anfig} and \ref{pofig}, respectively, whereas the
model parameter in the case (iii) is close to the border (b) in Fig.~\ref{validfig}.
In the cases (i), (ii), (iii) the effective gravitational couplings
in the regime $r_{2}\ll1$ are $G_{{\rm eff}}/G=1-0.05r_{2}$, $G_{{\rm eff}}/G=1+0.26r_{2}$,
$G_{{\rm eff}}/G=1+0.51r_{2}$, respectively, whereas at the de Sitter
solution $G_{{\rm eff}}/G=5.56$, $G_{{\rm eff}}/G=1.11$, $G_{{\rm eff}}/G=0.84$,
respectively. For smaller $\beta$, $G_{{\rm eff}}$ is larger in
the early epoch ($r_{2}\ll1$), so that the deviation of $\gamma$
from the value 0.55 is more significant. On the other hand, $G_{{\rm eff}}$
at the de Sitter solution gets smaller for decreasing $\beta$, which
leads to the approach to the value $0.55$ around $z=0$.

While the above discussion corresponds to the case $s=0.2$, we have
also studied the evolution of perturbations for different values of
$s$ and found the similar properties to those discussed above. For
$p=1$ the upper bound for the allowed parameter space in the $(\alpha,\beta)$
plane is characterized by the line $\beta=\alpha/2$ \cite{Defe2011},
around which the ISW-LSS anti-correlation is present for $0<s<0.36$
(i.e. for $s$ constrained observationally at the background level).
As $\beta$ gets smaller for a fixed $\alpha~(>0)$, the anti-correlation
tends to disappear. We also found that, for $p=1$ and $0<s<0.36$,
the growth index $\gamma$ varies in the range $0.4\lesssim\gamma\lesssim0.6$
for $0\le z\lesssim1$. The ISW-LSS anti-correlation as well the variation
of $\gamma$ allows us to discriminate between the extended Galileon
models and the $\Lambda$CDM.


\section{Conclusions}


In this paper we have studied cosmological constraints on the extended
Galileon models of dark energy. For the functions (\ref{geGali})
with the powers (\ref{power}) there exist tracker solutions along
which the field equation of state $w_{{\rm DE}}$ changes as $-1-4s/3$
(radiation era) $\to$ $-1-s$ (matter era) $\to$ $-1$ (de Sitter
era), where $s=p/(2q)>0$. Unlike the case of the covariant Galileon
($s=1$), $w_{{\rm DE}}$ can be close to $-1$ during the radiation
and matter eras for $0\leq s\ll1$. Moreover, even for $w_{{\rm DE}}<-1$,
there are viable model parameter spaces in which the ghosts and Laplacian
instabilities are absent.

Using the recent data of SN Ia, CMB, and BAO, we placed observational
constraints on the background tracker solutions with $s>0$. In the
flat Universe we found that the model parameters are constrained to
be $s=0.034_{-0.034}^{+0.327}$ and 
$\Omega_{m,0}=0.271_{-0.010}^{+0.024}$
(95\% CL) from the joint data analysis. The chi-square for the best-fit
case ($s=0.034$) is slightly smaller than that in $\Lambda$CDM.
However the difference of the AIC information criteria between the
two models is $\Delta({\rm AIC})=1.85$, so that the extended Galileon
is not particularly favored over the $\Lambda$CDM. We also carried
out the likelihood analysis in the presence of the cosmic curvature
$K$ and obtained the bounds $s=0.067_{-0.067}^{+0.333}$, $\Omega_{m,0}=0.277_{-0.022}^{+0.023}$,
and $\Omega_{K,0}=0.0077_{-0.0127}^{+0.0039}$ (95\,\%~CL). 
The tracker for the covariant Galileon ($s=1$) is disfavored from the data, in
which case only the late-time tracking solution is allowed observationally
\cite{Nesseris}.

The background quantities for the tracker are independent of the values
of $\alpha$ and $\beta$. This means that the models with different
$\alpha$ and $\beta$ cannot be distinguished from the observational
constraints derived from the background cosmic expansion history.
In order to break this degeneracy we studied the evolution of cosmological
perturbations in the presence of non-relativistic matter for the flat
FLRW background. As shown in Ref.~\cite{Kimura}, for $\alpha=\beta=0$,
the matter density perturbation $\delta_{m}$ divided by the scale
factor $a$ is anti-correlated with the effective gravitational potential
$\Phi_{{\rm eff}}$ for the modes relevant to the LSS. This leads
to the anti-correlation between the LSS and the ISW effect in CMB,
so that the parameter $s$ for the tracker is severely constrained
to be $s<1.2\times10^{-4}$ (95\,\% CL).

For the models with $\alpha\neq0$ and $\beta\neq0$, however, the
correlation between $\delta_{m}/a$ and $\Phi_{{\rm eff}}$ depends
on the values of $\alpha$ and $\beta$. If $p=1$ and $q=5/2$ (i.e.
$s=0.2$), for example, $\delta_{m}/a$ and $\Phi_{{\rm eff}}$ tend
to be positively correlated for the model parameters close to the
border (b) in Fig.~\ref{validfig} ($\beta=\alpha/2-1/15$), whereas
they show anti-correlations for $\alpha$ and $\beta$ close to the
border (a) ($\beta=\alpha/2$). The typical examples of the positive
and negative correlations are plotted in Figs.~\ref{pofig} and \ref{anfig},
respectively. The qualitative differences between these two cases
can be understood by estimating the effective gravitational coupling
$G_{{\rm eff}}$ and the quantity $\xi=(G_{{\rm eff}}/G)(1+\eta)/2$
derived under the quasi-static approximation on sub-horizon scales.
As the model parameters approach the border (a) in Fig.~\ref{validfig},
$\xi$ gets larger while $G_{{\rm eff}}$ decreases, so that the anti-correlation
between $\delta_{m}/a$ and $\Phi_{{\rm eff}}$ tends to be stronger.
We studied the evolution of perturbations for different values of
$s$ in the range $0<s<0.36$ and found that the basic properties
for the ISW-LSS correlation are similar to those discussed for $s=0.2$.

We also estimated the growth index $\gamma$ of the matter perturbation
and found that, for $p=1$ and $0<s<0.36$, it typically varies in
the range $0.4\lesssim\gamma\lesssim0.6$ at the redshifts for $0\le z\lesssim1$.
Hence it is also possible to distinguish between the extended Galileon
models and the $\Lambda$CDM model from the galaxy clustering. However,
we expect that the tightest observational bounds on the values $\alpha$
and $\beta$ should come from the ISW-LSS correlation. We leave such
observational constraints for future works.


\section*{ACKNOWLEDGEMENTS}

\label{acknow} We thank Savvas Nesseris for providing us a related
numerical code for the likelihood analysis of CMB, BAO, and SN Ia.
The work of A.\,D.\,F.\ and S.\,T.\ was supported by the Grant-in-Aid
for Scientific Research Fund of the JSPS Nos.~10271 and 30318802.
S.\,T.\ also thanks financial support for the Grant-in-Aid for Scientific
Research on Innovative Areas (No.~21111006). 



\end{document}